\newif\ifFull
\Fulltrue
\documentclass[12pt]{article}
\usepackage{amsmath, amssymb, algorithm, algorithmic, color}
\usepackage{enumerate, graphicx, hyperref}
\usepackage{cite}
\usepackage{wrapfig}
\ifFull\else
\usepackage{mathptm}
\fi

\ifFull\else
\let\doendproof\endproof
\renewcommand\endproof{~\hfill\qed\doendproof}

\fi

\ifFull\else
\DeclareSymbolFont{largesymbols}{OMX}{cmex}{m}{n}
\fi

\ifFull\else
\fi
\title{Force-Directed Graph Drawing \\ Using Social Gravity and Scaling}

\ifFull
\author{Michael J. Bannister
\and David Eppstein
\and Michael T. Goodrich
\and Lowell Trott \and \\
Department of Computer Science, University of California, Irvine}
\else
\author{Michael J. Bannister
\and David Eppstein
\and Michael T. Goodrich
\and Lowell Trott}
\institute{Department of Computer Science, University of California, Irvine}
\fi

\begin{document}
\maketitle

\begin{abstract}
Force-directed layout algorithms produce graph drawings by resolving 
a system of emulated physical forces.  
We present techniques for using \emph{social gravity} as an additional
force in force-directed layouts, together with a \emph{scaling} technique,
to produce drawings of trees and forests, 
as well as more complex social networks.
Social gravity assigns mass to vertices in proportion
to their network centrality, which allows vertices that are more graph-theoretically central 
to be visualized in physically central locations.  Scaling varies the gravitational force throughout the simulation, and reduces crossings relative to unscaled gravity.
In addition to providing this algorithmic framework,
we apply our algorithms to social networks produced by Mark Lombardi, 
and we show how social gravity can be incorporated 
into force-directed Lombardi-style drawings.
\end{abstract}

\section{Introduction}
In a \emph{social network}, vertices 
represent social entities, such as people or corporations, and
edges represent relationships, such as friendship or partnership,
between pairs of such entities.
Online social networks, such as Facebook and Google+, are common illustrations of people as actors, and mutual friendship as edge relations~\cite{Fur-10}.
The study of social networks, therefore,
provides a methodological approach to understanding 
social structure using graph-theoretic techniques 
(e.g., see~\cite{BorMehBra-09}).  

In the study of social networks, it is often important to distinguish
between the importance or \emph{centrality} of its vertices, or more accurately,
of the entities they represent.
Social networking researchers have identified several different centrality measures, each of which provides a way 
to relate the value of a social entity
to the structure of the network to which it belongs.  
We consider three of the most commonly-used types of centrality in this paper: 
\emph{degree centrality}, \emph{betweenness centrality},
 and \emph{closeness centrality} (e.g., see~\cite{Fre-SN-78}). 
The degree centrality of vertex $v$ is defined simply to be its degree. 
The closeness centrality of a vertex $v$ is defined to be 
the reciprocal of the mean distance from $v$ to all other vertices in 
the network. 
The betweenness centrality of a vertex $v$ is defined to be the 
proportion of paths containing $v$ among all shortest paths in the network, which can be expressed as the sum
\[
\sum_{s \neq t \neq v} {\sigma_{st}(v)}/{\sigma_{st}},
\]
where $\sigma_{st}$ is the number of shortest paths from $s$ to $t$, 
and $\sigma_{st}(v)$ is the number of shortest paths from $s$ to $t$ 
that contain $v$.  
For more information on centrality and its efficient computation, 
see~\cite{SteZel-SN-89, Bra-JMS-01}.

Naturally, the
drawing of social networks is an important part of social science research
(e.g., see Brandes {\it et al.}~\cite{bfw-sn-13}).
Since centrality is a common focus of social network analysis, 
we feel that
placing vertices with significant centrality close to the 
middle of a drawing should be a natural 
goal of drawing algorithms for social networks, and the approach we
explore in this paper for achieving this goal
is based on the framework of force-directed layouts.

Force-directed layout algorithms are well-known in 
the graph drawing literature, as they yield reasonable drawings for 
a wide variety of graphs
(e.g., see~\cite{BraHimRoh-GD-96, Bra-DG-05, BatEadTam-98, FruRei-SPE-91,%
GajGooKob-CG-04, GajKob-GD-01}).
Traditionally, these methods use a graph's structure to mimic a 
physical system of attractive spring forces along edges and 
universal repulsive forces emanating from vertices.  
The system of resulting force equations is then resolved into a 
minimal-energy state, which produces a drawing.  

Frick {\it et al.}~\cite{FriLudMeh-GD-94}
introduce the use
of gravitational forces in such layout methods
in addition to spring and repulsive forces,
and similar gravitational forces have been utilized in several
other force-directed systems as well
(e.g., see~\cite{de-vfmfg-02,Sander1999175}).
The forces used in these systems are
intended to pull vertices towards
the center of a drawing using a ``mass''
that is proportional to vertex degree (that is, 
using something akin to degree centrality), 
and, in so doing, to speed up algorithmic convergence.
In this paper, we study force-directed
methods based on the use of a generalized gravitational force,
which we call \emph{social gravity}, where
the mass of a vertex
can depend on any social-networking centrality measure.

\subsection{Related Work}
Brandes {\it et al.}~\cite{BraKenRaa-M-06} 
describe a network visualization tool that
places 
vertices on concentric rings whose radii are based on centrality measures.
Correa {\it et al.}~\cite{ccm-vrsn-12}
perform a study
based on visual reasoning about social networks and about how centrality can be
used as an aid in network simplification, including use of a tool that
places
vertices on concentric rings in a manner similar
that that of Brandes {\it et al.}~\cite{BraKenRaa-M-06}.
Brandes and Pich~\cite{BraPic-JGAA-11} 
show how to maintain the approach of such radial layouts while using stress
minimization.
These works differ from our use of social gravity in that 
radial drawings constrain vertices to discrete rings based on 
centrality values, whereas force-directed placement using social gravity is 
more continuous.
\ifFull
For additional work on other techniques for 
visualizing social networks, please see, 
e.g.,~\cite{BraKenRaa-JTP-99, BraKenRaa-M-06, Fre-JSS-00, Fre-MMSNA-05,%
Klo-SN-81}.
\fi

A prime example of culturally relevant social network drawing 
is the work of Mark Lombardi~\cite{lh-mlgn-03}.  
Mark Lombardi was an American artist who created 
drawings attempting to document financial and political conspiracies.  
His drawings are recognizable for the use of circular arcs for edges and for their aesthetically pleasing placement of vertices and edges.  
Lombardi's art has inspired algorithmic methods in graph drawing
(e.g., see~\cite{degkl-ppld-12,degkn-dtwpa-10,degkn-ldg-10}), combining the ideas of drawing edges as circular arcs and of optimizing the angular resolution of the vertices.
Most relevantly for our work, Chernobelskiy {\it et al.}~\cite{ccgkt-fdlgd-12}
introduce a force-directed approach to produce 
Lombardi-style graph drawings 
with perfect or near-perfect angular resolution and circular-arc edges.
They do not consider gravitational forces in their methods, however.

\subsection{Our Results}
As mentioned above,
force-directed algorithms are an efficient and effective approach 
to graph drawing.  
In this paper, we study methods for harnessing 
\emph{social gravity} to control the positioning of vertices in 
a force-directed layout.
That is, we show how to incorporate gravitational forces into a
force-directed layout so that
vertices are affected by gravity in relation to various
social-networking centrality measures.  

As it is in the physical world, gravity
can be a powerful force in graph drawing, which can overwhelm other forces. In force-directed algorithms, strong gravitational forces can cause a drawing to ``freeze'' into a local minimum configuration with many crossings. To ameliorate this effect, we use gravity in conjunction with a
\emph{scaling} technique: we perform a force-directed simulation that starts without gravity,
so as to untangle edges and reduce crossings, and 
we then gradually increase the influence of gravity so as to push 
vertices to the center in proportion to their importance.
In contrast, previous force-directed methods using gravity~\cite{FriLudMeh-GD-94}
have a parameter similar to our variable scaling factor, $\gamma_t$, 
but set it to be a fixed constant.

Using social gravity and scaling
allows us to produce drawings that have all the 
classic advantages of force-directed algorithms, such as being
space efficient and visually pleasing, while also 
centralizing socially significant vertices.  
We examine the social networks of Mark Lombardi as a test case that shows that combining our ideas of social gravity and scaling in conjunction with 
Lombardi-style force-directed drawing~\cite{ccgkt-fdlgd-12} can 
create effective graph drawings for social networks.

\section{The Algorithm}
Our algorithm has a structure common to most force-directed algorithms. For
the forces considered, we attempt to place vertices of the graph such that they
are in equilibrium, i.e., so that the net force on each vertex is zero. To achieve this
goal we compute the net force on each vertex, and then move the vertex in the
direction of this net force. 
Convergence is sped up by normalizing the net force down. 
This two-step process is then repeated until the vertices are approximately in equilibrium or the maximum number of iterations has been reached.

The forces we consider in our algorithm are the attractive and repulsive  forces of Fruchterman and Reingold~\cite{FruRei-SPE-91},
\[
f_r(u,v) = \frac{k^2}{\Vert P[u] - P[v] \Vert^2}(P[v] - P[u]) \quad f_a(u,v)= \frac{\Vert P[u] - P[v]\Vert}{k}(P[u] - P[v])
\]
where $P[v]$ is the current position of vertex $v$, and $k$ is a parameter controlling the \emph{natural length} of an edge so that, e.g., when drawing $K_2$ its edge will have length $k$. We then augment these forces with a gravitational force
\[
f_g(v) = \gamma_t M[v](\xi - P[v])
\]
where $M[v]$ is the mass of vertex $v$, $\xi = \sum_v P[v] / \vert V \vert$ is the centroid of the points, and $\gamma_t$ is the gravitational scaling parameter that varies per iteration. This gravitational force is inspired by the work of Frick {\it et al.}~\cite{FriLudMeh-GD-94}. However, as mentioned above, we consider mass based on social networking parameters as opposed to just degree. In addition, we consider the gravitational force primarily as a force for shaping the drawing rather than as a tool to speed up convergence, so we use larger values for $\gamma_t$ than Frick {\it et al.}

To reduce crossings in the final drawing, our \emph{scaling} technique gradually increases the value for $\gamma_t$ through the iterations.
For the drawings in this paper, we control this increase using a step function, where $\gamma_t$ starts at $0$ and increases by approximately $0.2$ when the vertices reach an approximate equilibrium. 
How high $\gamma_t$ should go may depend on the graph, but a maximum value of 2.5 seems to yield pleasing results. 

\subsection{Pseudo-code}
Our implementation starts by precomputing the mass of each vertex and storing these values in a dictionary $M$. 
The computation of $M$ varies depending on which form of centrality is being used.
After this initialization we run the Algorithm~\ref{alg:main}, shown below.

\begin{algorithm}[b!]
\caption{Force-directed layout with social gravity and scaling}
\label{alg:main}
\begin{algorithmic}
\FOR {$t\gets 1$ to {\tt max\_iteration}}
  \STATE $\gamma_t \gets$ according to a scaling schedule
  \FOR {each vertex $v$ in graph $G$}
    \STATE $\displaystyle I[v] \gets \sum_{u \neq v} f_r(u,v) + \sum_{\{u,v\} \in E} f_a(u,v) + \sum_v f_g(v)$
  \ENDFOR
  \FOR {each vertex $v$ in graph $G$}
    \STATE $P[v] \gets P[v] + \sigma \cdot \min(I_\text{max}, I[v])$
  \ENDFOR
\ENDFOR
\end{algorithmic}
\end{algorithm}

Before each iteration, the gravitational parameter, $\gamma_t$, 
is updated according to a scaling schedule.
After the impulse force is computed for each vertex, 
it is capped to $I_\text{max}$ and scaled by $\sigma$. 
In our runs, we typically set $k = 80$, $I_\text{max} = 10$, 
$\sigma = 0.1$, $\gamma_t = 0.2\lfloor t / 200 \rfloor$, 
and \verb+max_iteration+ is set so that $\gamma_t \le 2.5$.
Example applications of our algorithm are given 
in Sections~\ref{sec:tree}~and~\ref{sec:soc-net}.

\section{Trees and Forests}\label{sec:tree}

\begin{figure}[hbt!]
\begin{center}
\includegraphics[trim = 70mm 0mm 0mm 0mm ,clip,width=.3\textwidth]{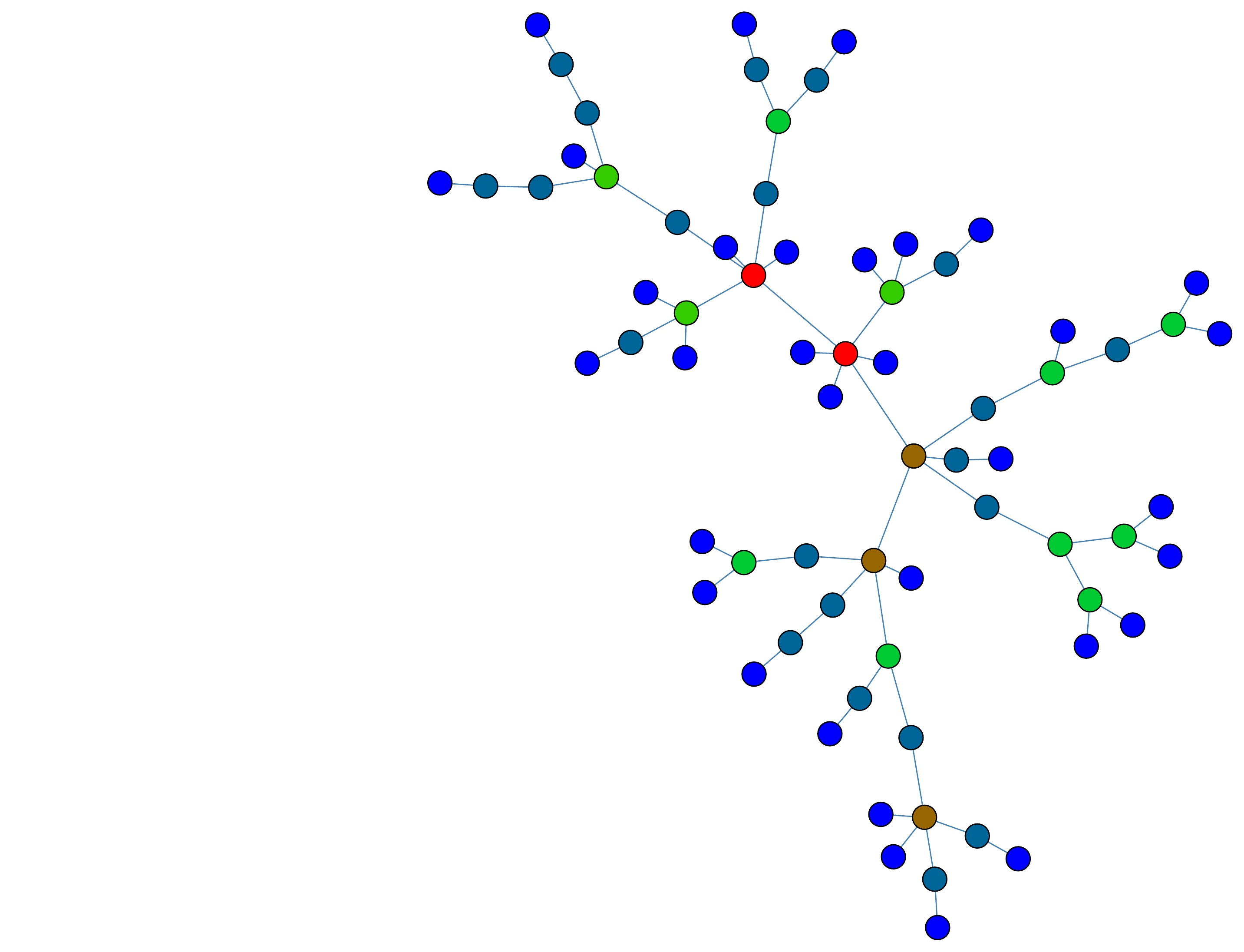}
\quad
\includegraphics[trim = 70mm 0mm 0mm 0mm ,clip,width=.3\textwidth]{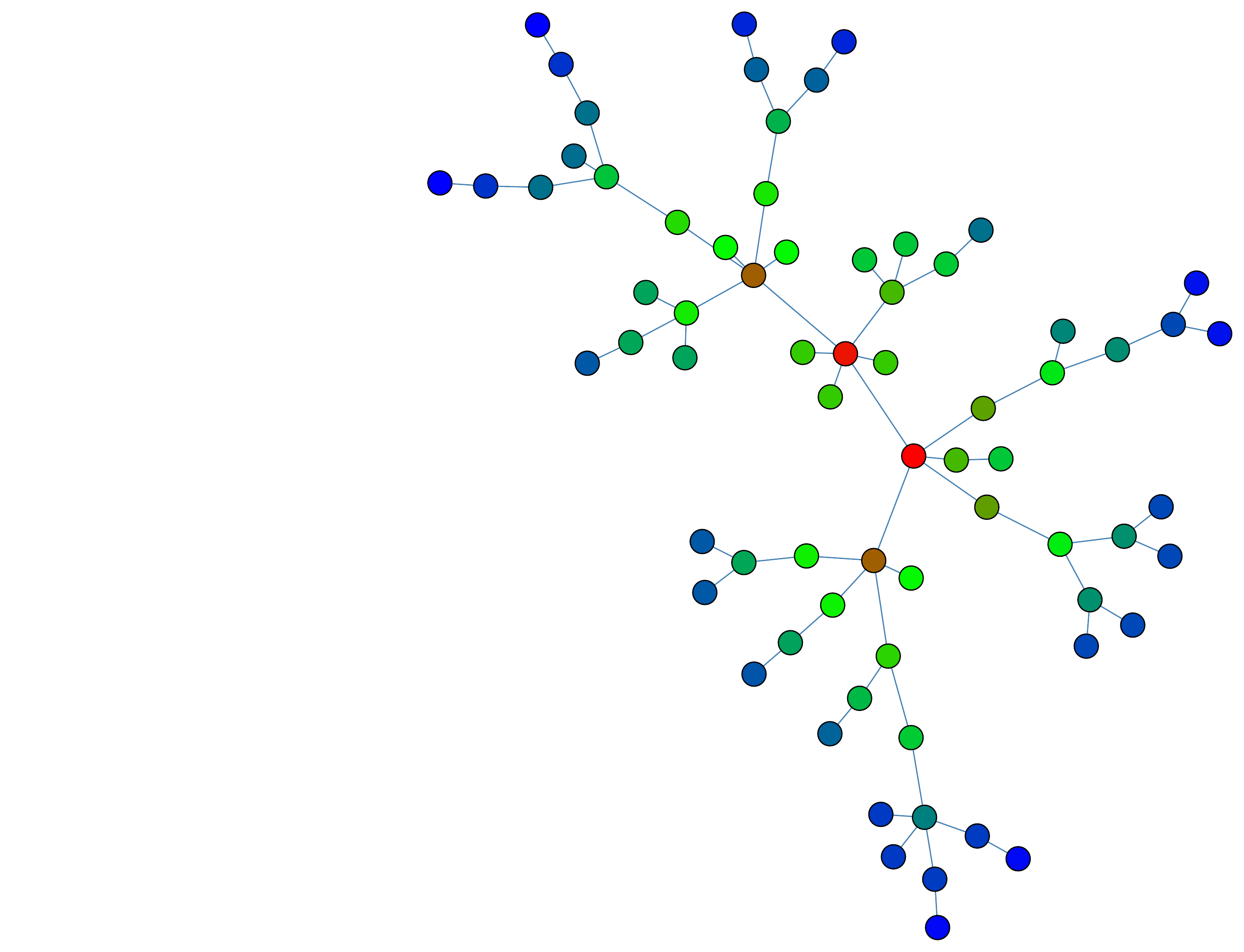}
\quad
\includegraphics[trim = 70mm 0mm 0mm 0mm ,clip,width=.3\textwidth]{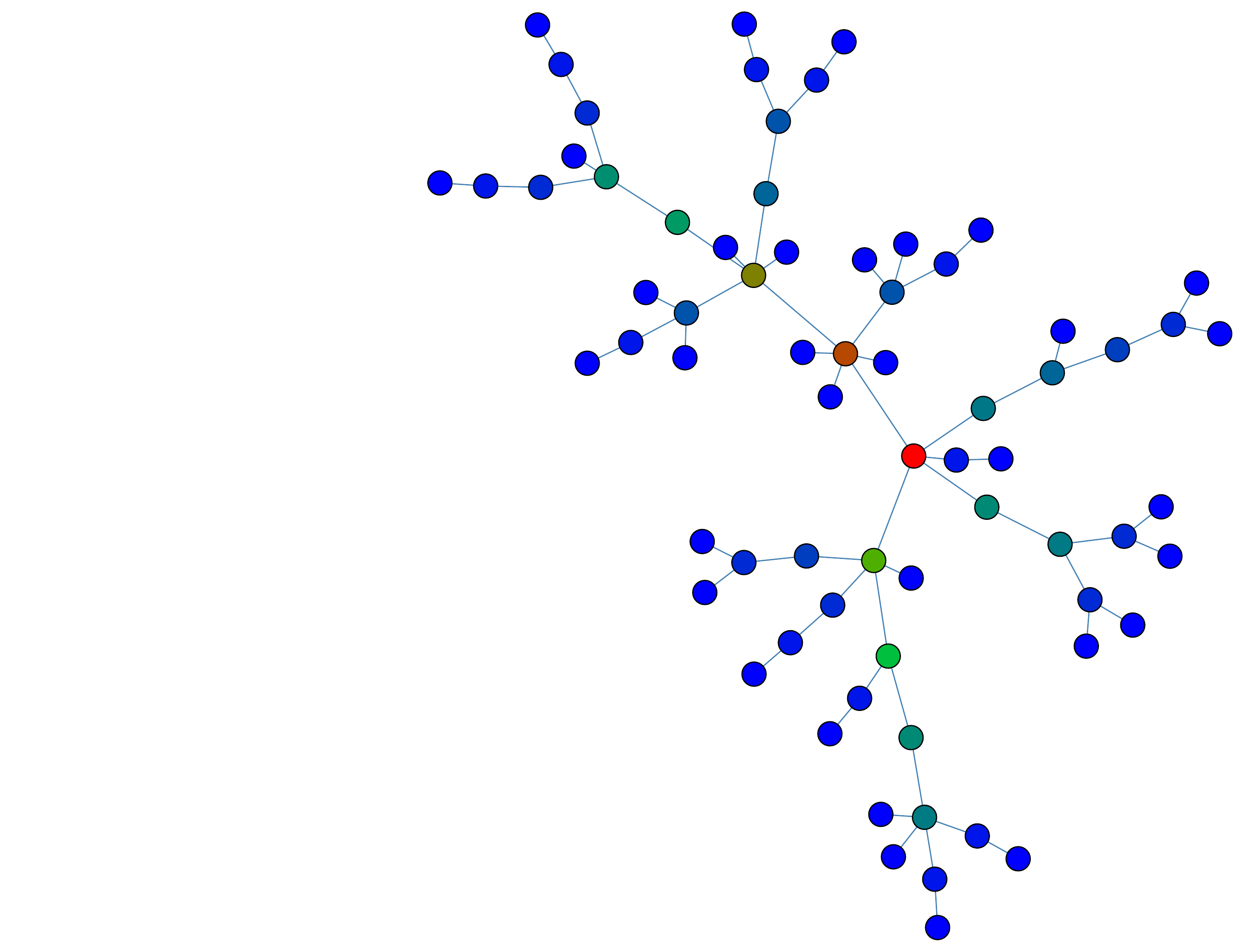}

\vspace{2em}

\includegraphics[width=.3\textwidth]{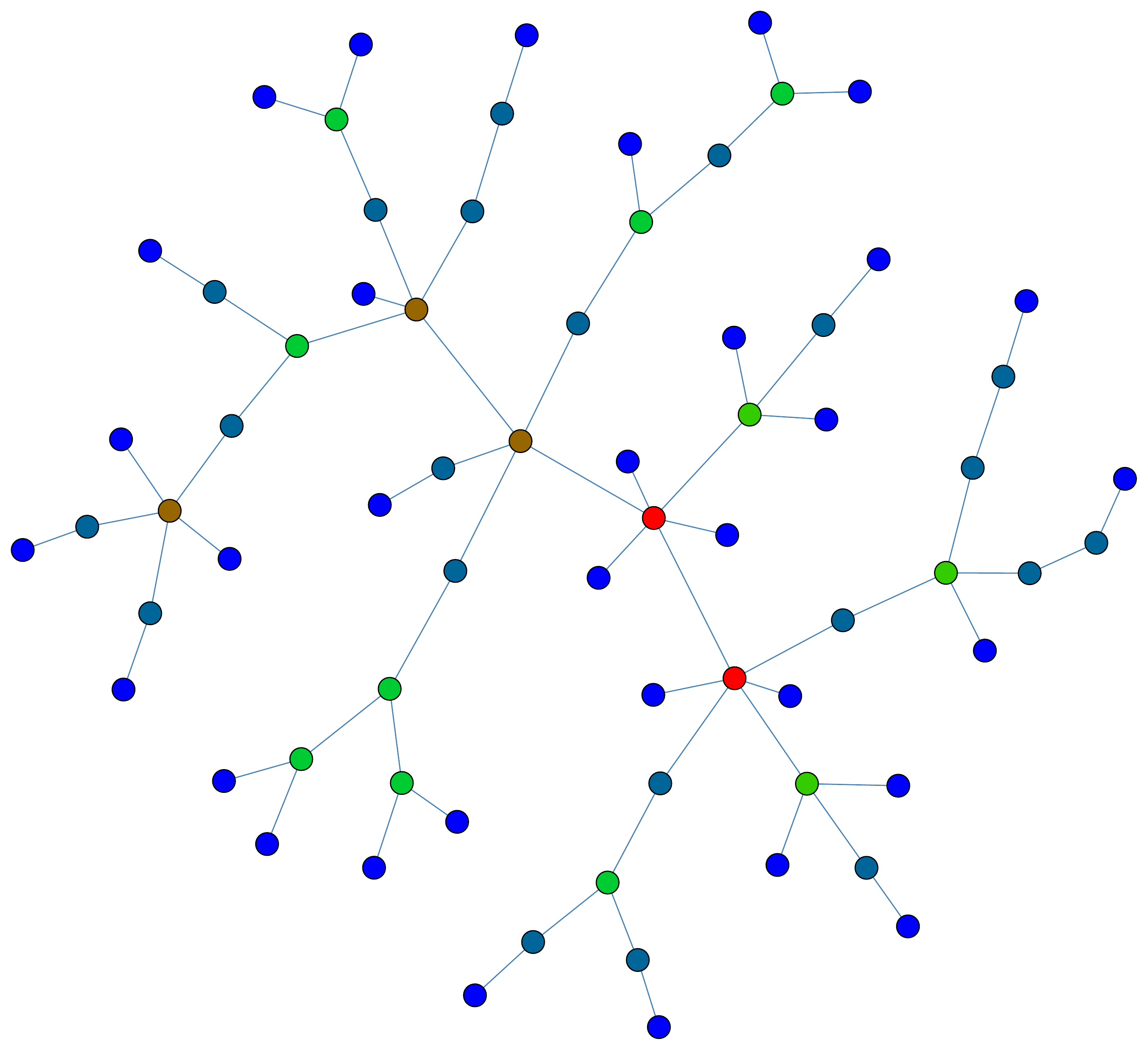}
\quad
\includegraphics[width=.3\textwidth]{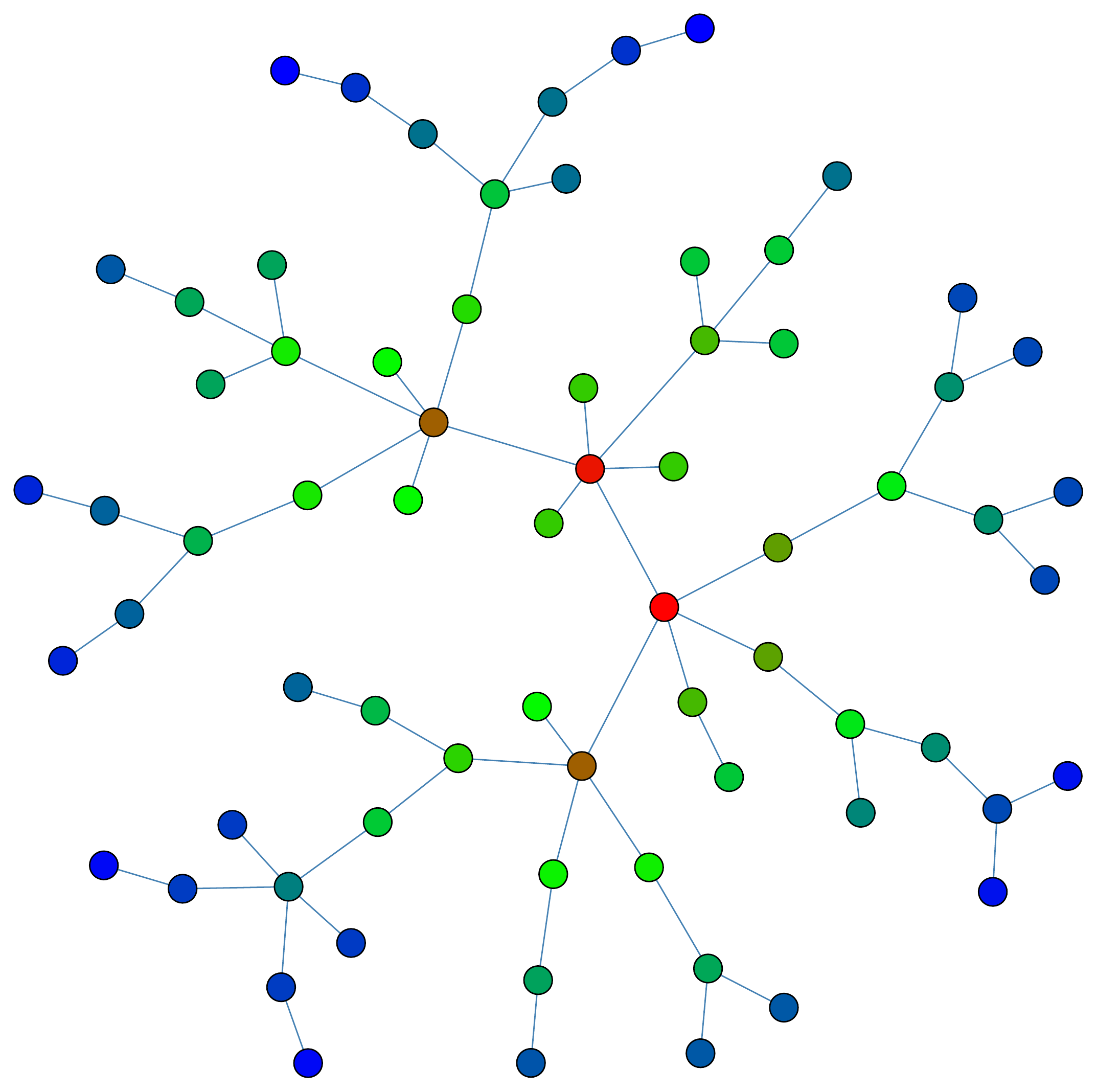}
\quad
\includegraphics[width=.3\textwidth]{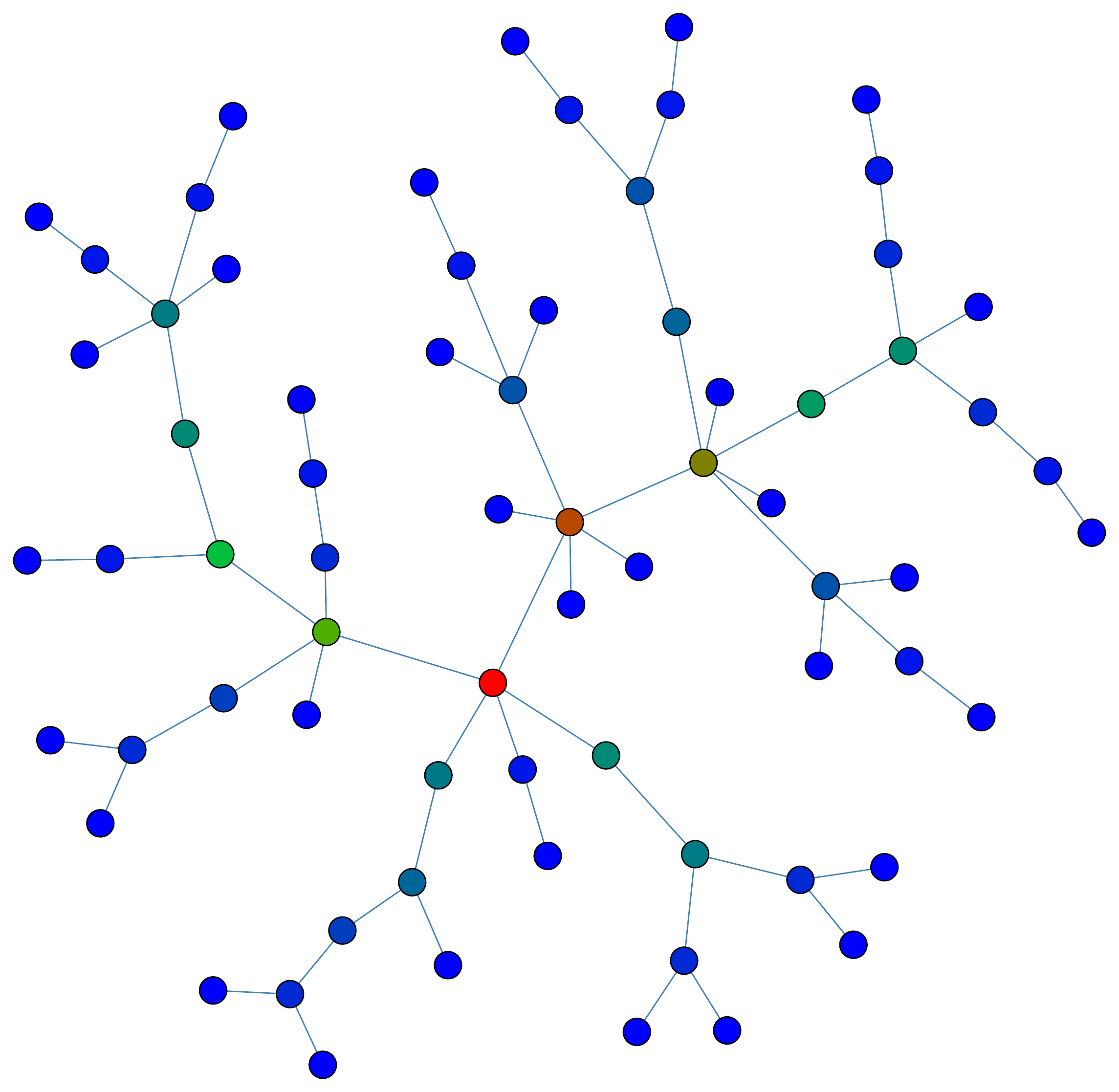}

\end{center}
\caption{\small\sf A tree of 70 vertices.  The top row is the classic force-directed embedding colored from left to right with degree, closeness, and betweenness centrality respectively.  The bottom row has the corresponding gravitational force-directed embedding. }
\label{fig-t70}
\end{figure}

As a first demonstration of our algorithm, we show how it performs on trees
and forests. 
The algorithm takes as input an unrooted tree, 
and makes no explicit attempt to identify a root, but our drawings can be viewed as being rooted at a node near the center of the drawing, which often tends to be one of high degree, closeness, or betweenness centrality. 
In Figure~\ref{fig-t70}, we see that our algorithm (bottom row) places the vertices of the graph uniformly in a disk, whereas the standard force-directed algorithms place the graph in a more elongated region. 
In addition to better space utilization, our algorithm tends to have better angular resolution than existing force directed algorithms, as can also be seen in
Figure~\ref{fig-t70}. 

\begin{figure}[hbt!]
\vspace*{-24pt}
\centering
\includegraphics[width=0.55\textwidth]{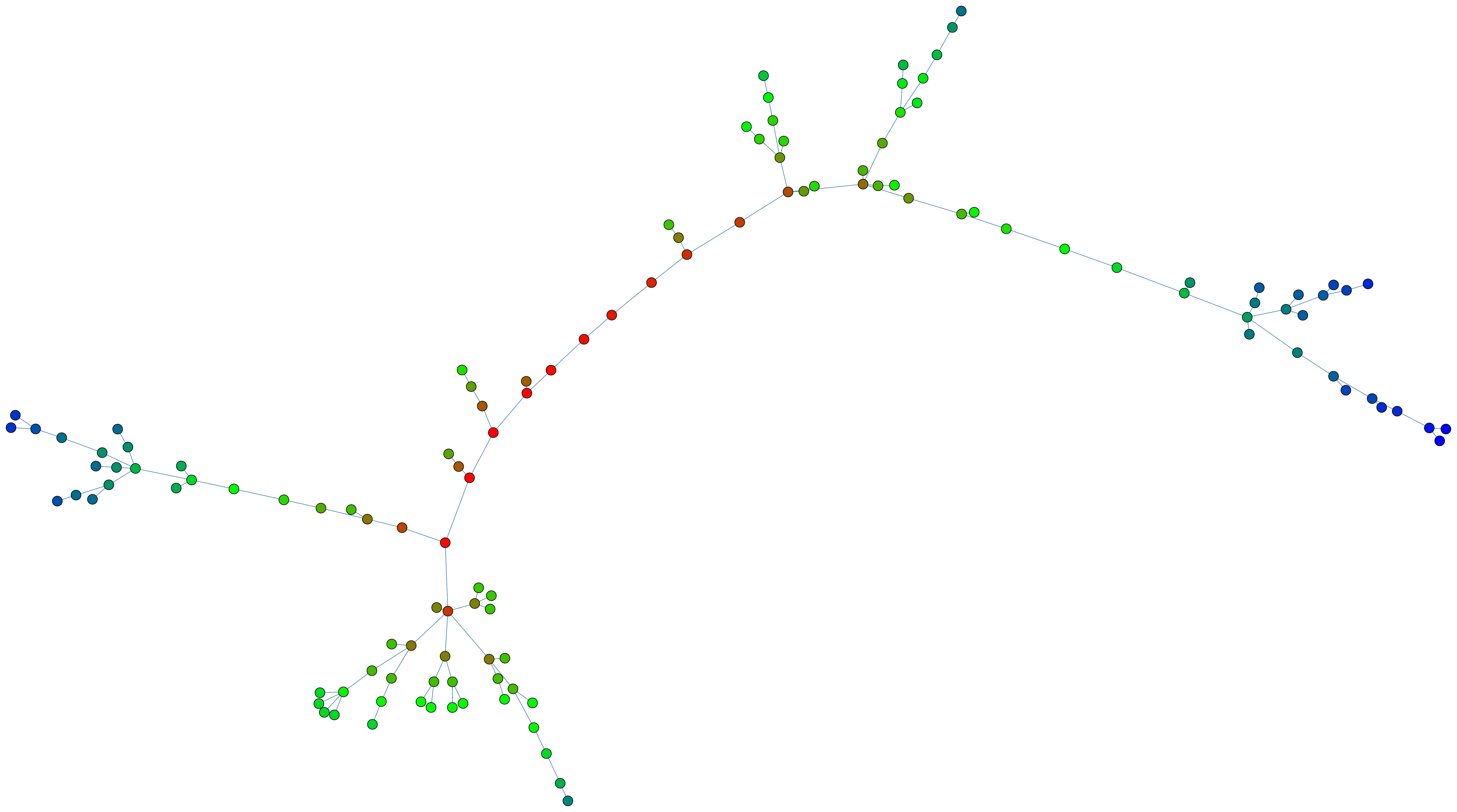}\quad
\includegraphics[width=0.35\textwidth]{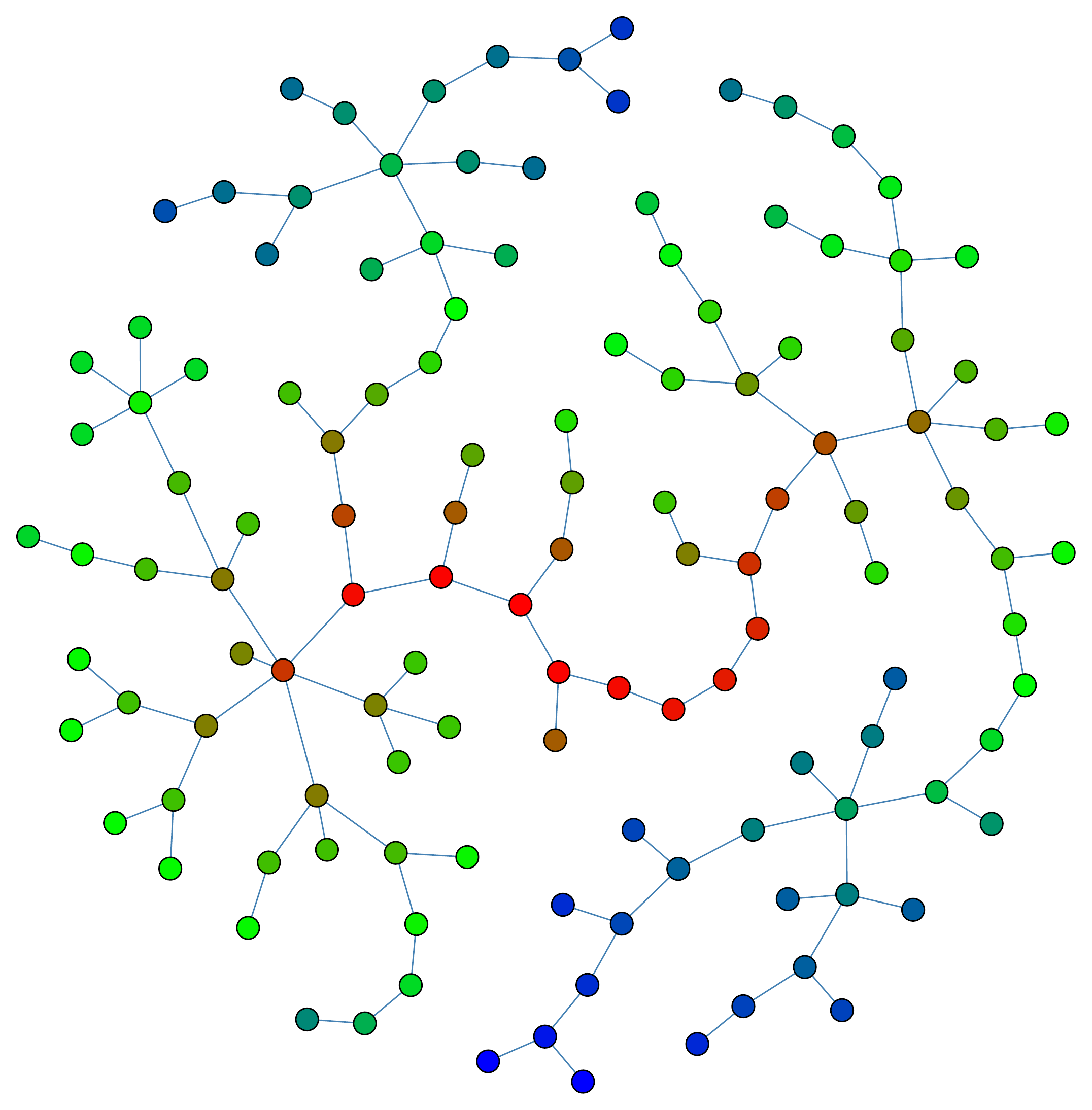}
\caption{\small\sf A tree with 126 vertices drawn on the left with the classical force based algorithm, and drawing with scaled gravity on the right.}
\label{fig-t126}
\end{figure}

\begin{figure}[hbt!]
\vspace*{-6pt}
\centering
\includegraphics[width=0.55\textwidth]{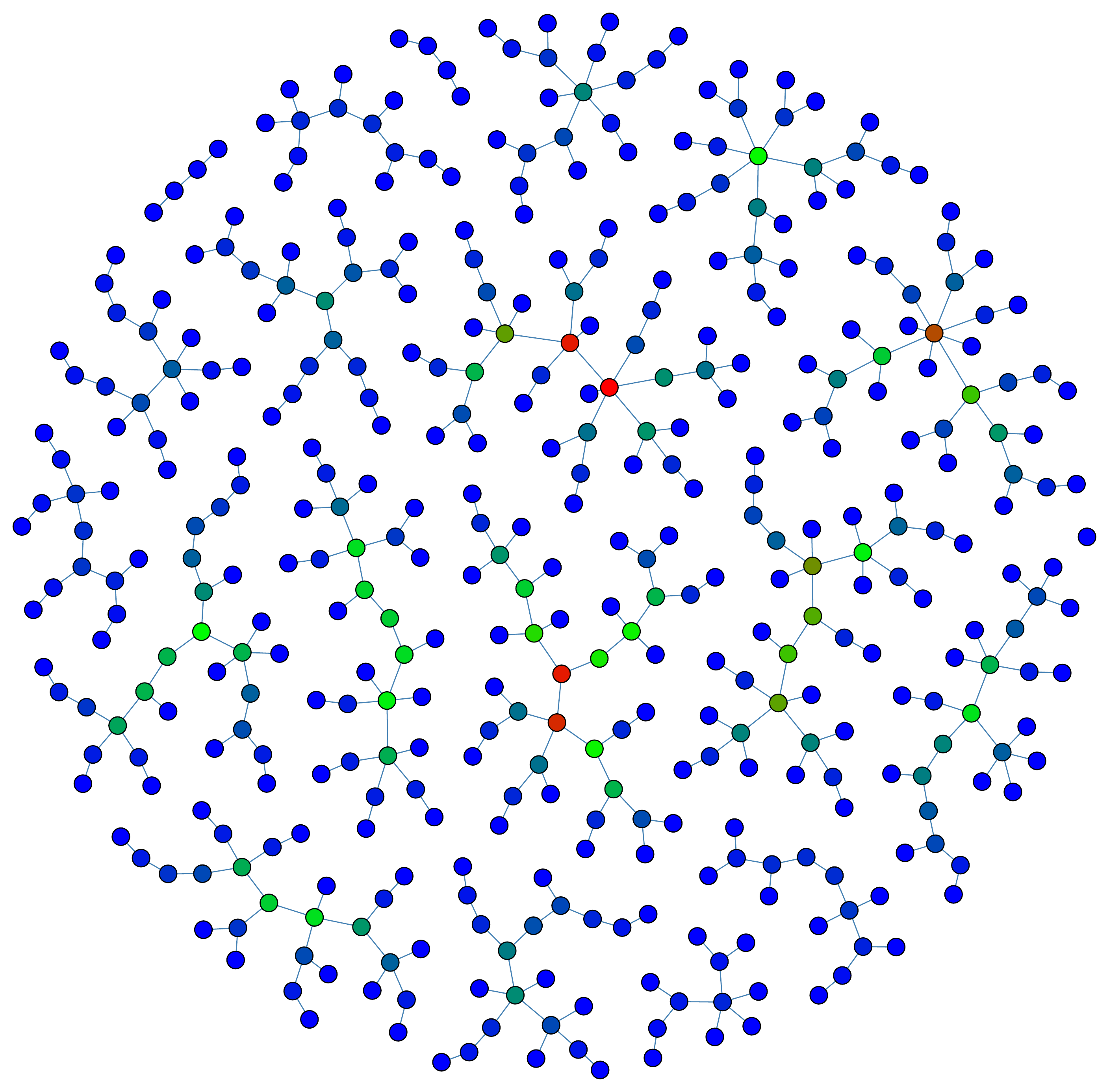}
\vspace*{-6pt}
\caption{\small\sf A forest with 422 vertices and 20 trees, drawn with gravity based on betweenness centrality.}
\label{fig-f20_40}
\end{figure}

An added benefit of the social-gravity force is that it 
allows a force-directed algorithm to draw trees 
more compactly, which improves the area bound for a drawing and 
allows for individual vertices to be drawn larger than they would 
be in the more widely-spaced drawings produced by traditional 
force-directed methods.
This effect becomes even more noticeable as the size of the graph increases, 
as can be see in Figure~\ref{fig-t126}. 
This compaction is done without introducing clutter or many crossings, 
due to our use of scaling to initially emphasize the
repulsive force that applies to vertices.
The compacting effect of social gravity
is especially helpful for drawings of disconnected forests, since
it keeps the trees that form the forest
from drifting apart, as shown in Figure~\ref{fig-f20_40}. 

\begin{figure}[hbt!]
\begin{center}
\includegraphics[width=.3\textwidth]{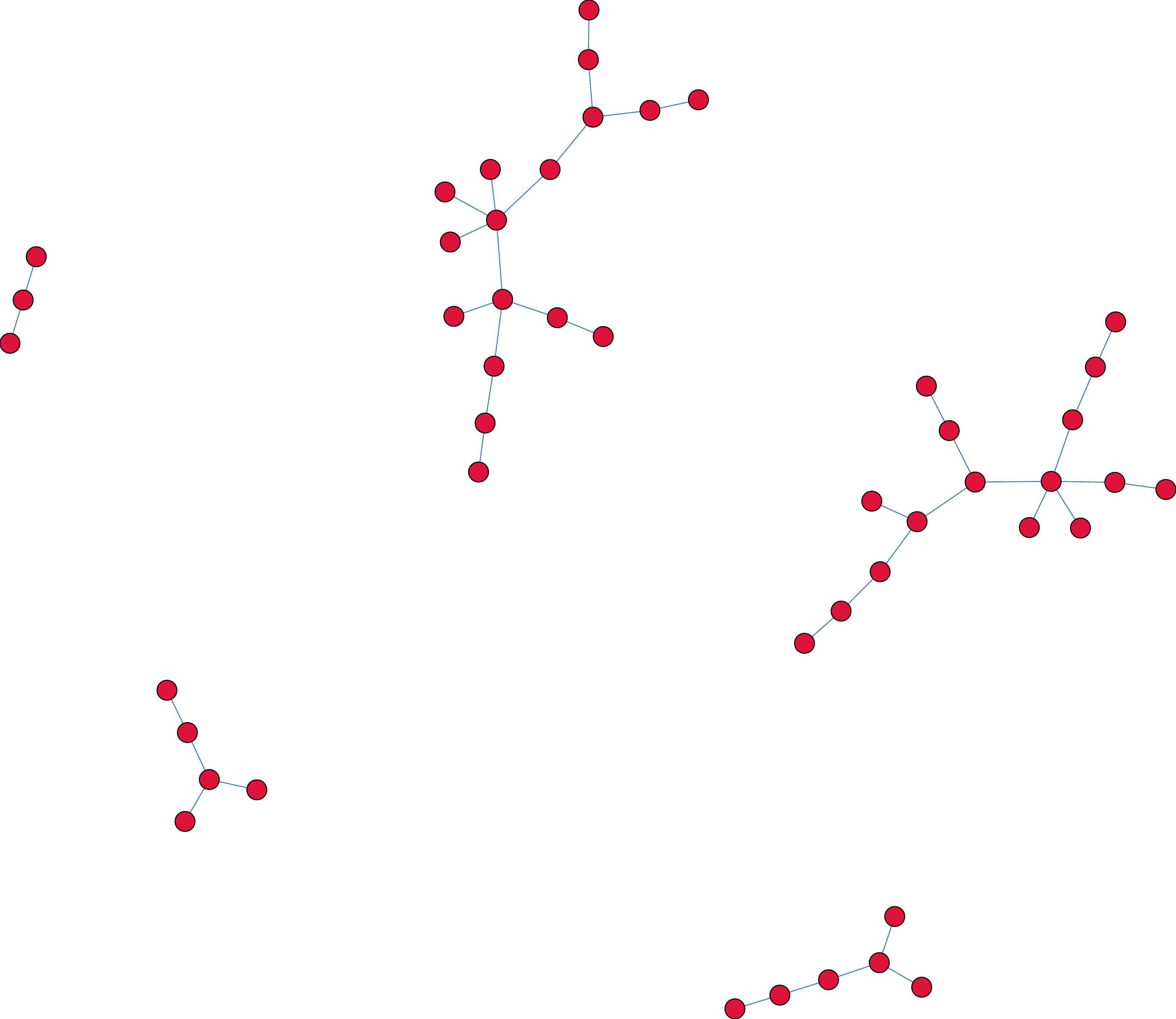}
\quad
\includegraphics[width=.3\textwidth]{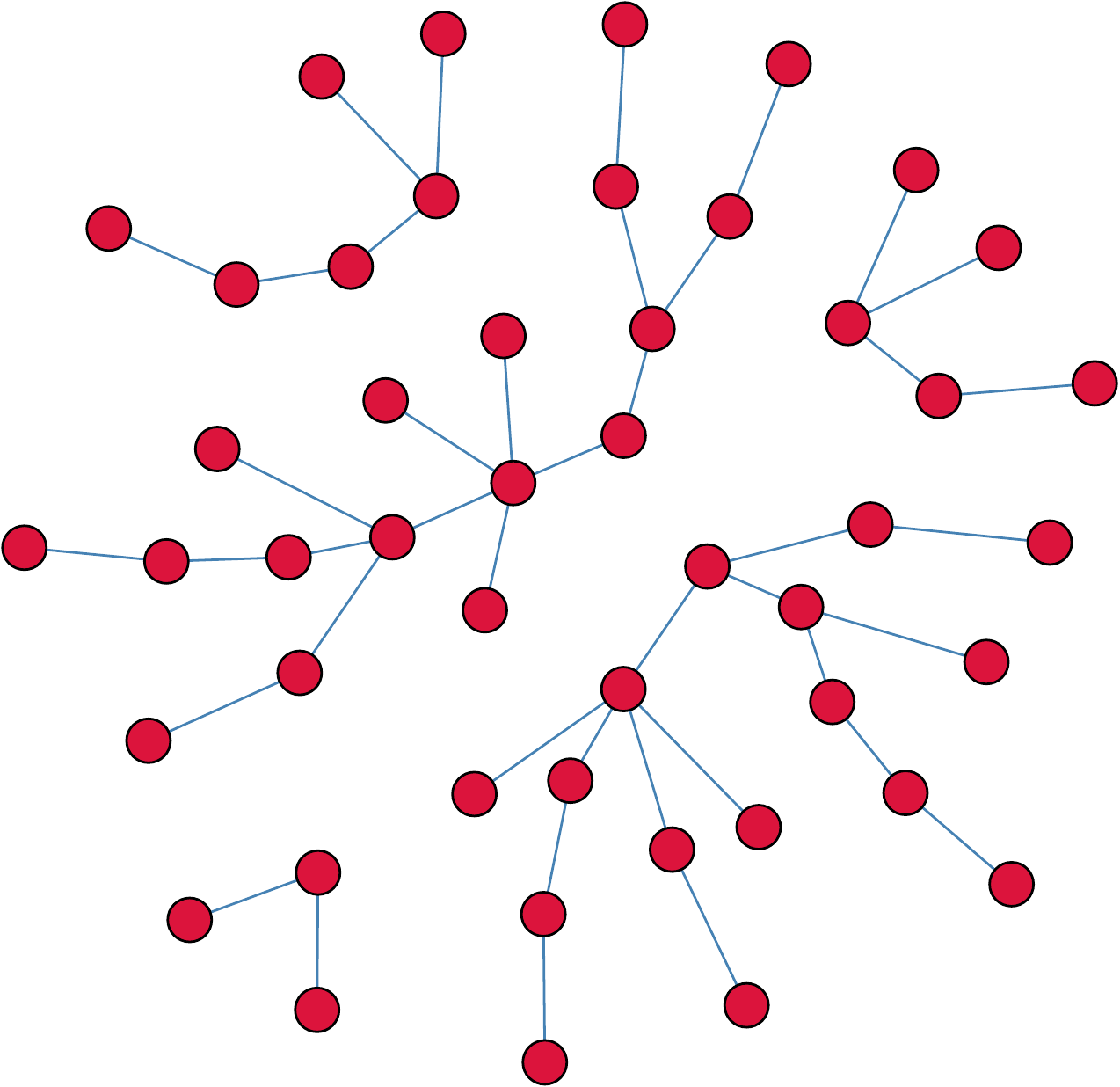}
\quad
\includegraphics[width=.3\textwidth]{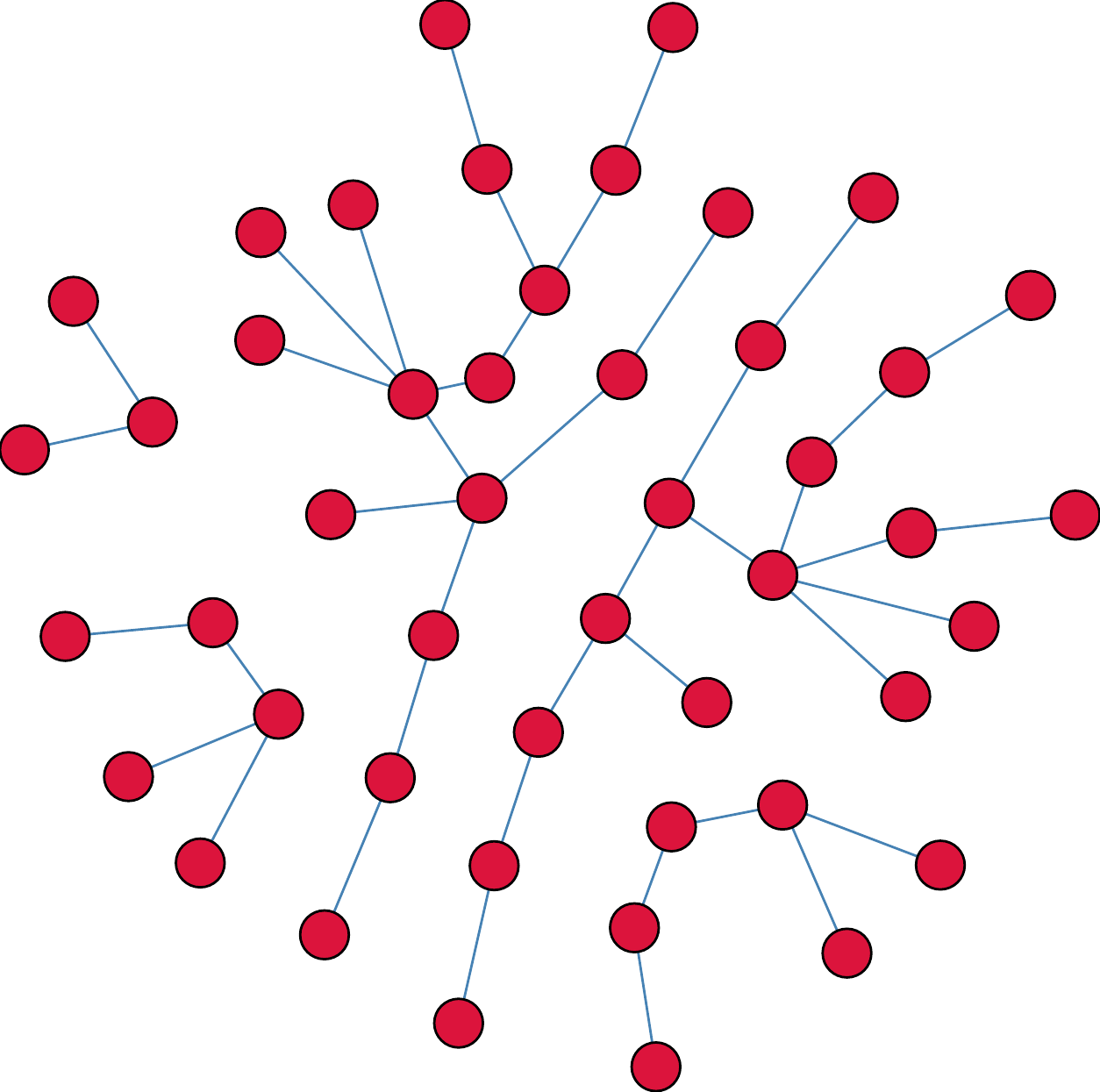}
\end{center}
\caption{\small\sf A five component forest with 45 vertices drawn without and with a gravitational force.  
The figure on the left has no gravity, in the middle gravity is scaled up, while the figure on the right has uniformly strong gravity throughout the algorithm. }
\label{fig-forest5-20-degGrav}
\end{figure}

\begin{figure}[hbt!]
\begin{center}
\includegraphics[width=.4\textwidth]{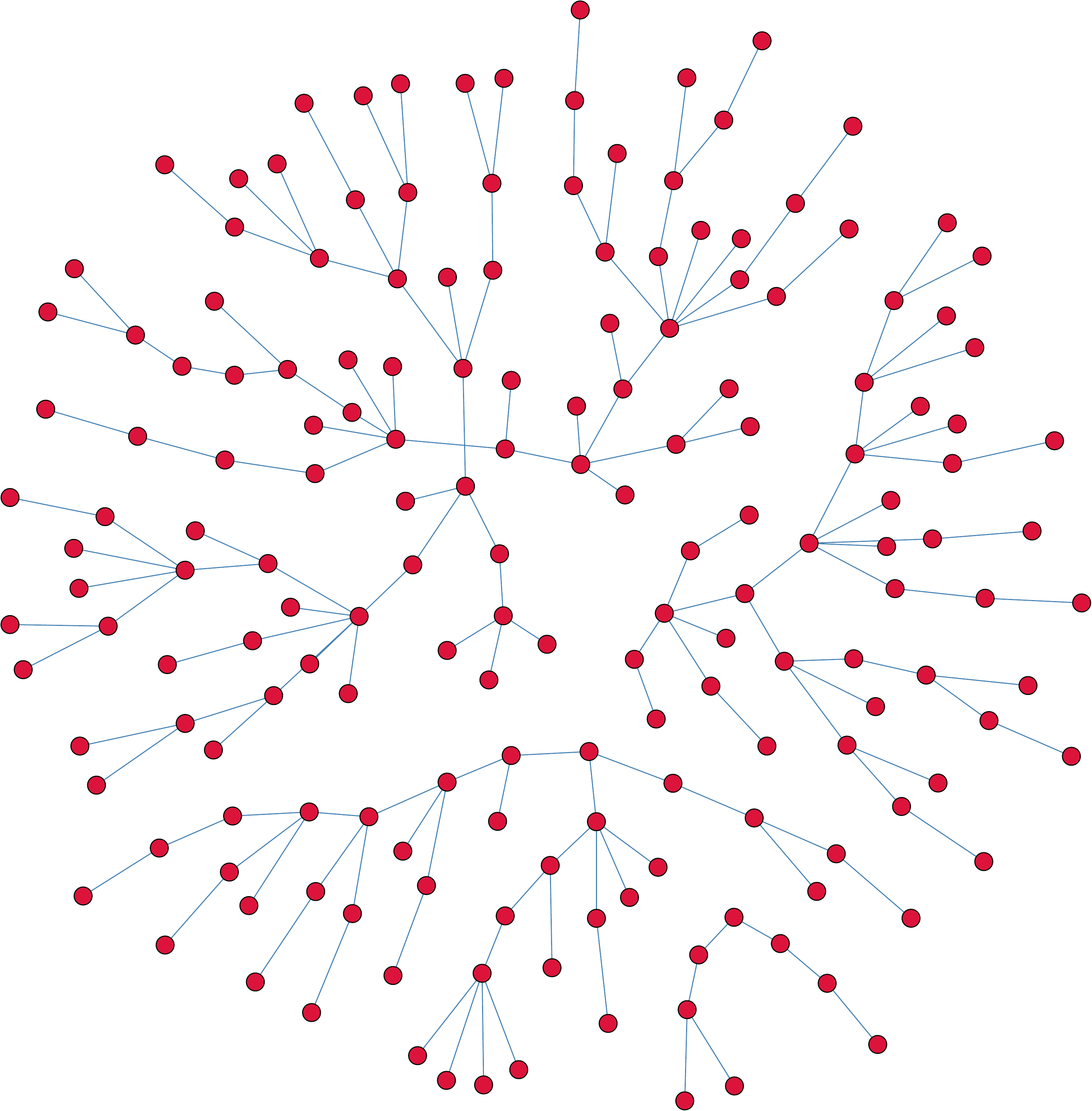}
\qquad
\includegraphics[width=.4\textwidth]{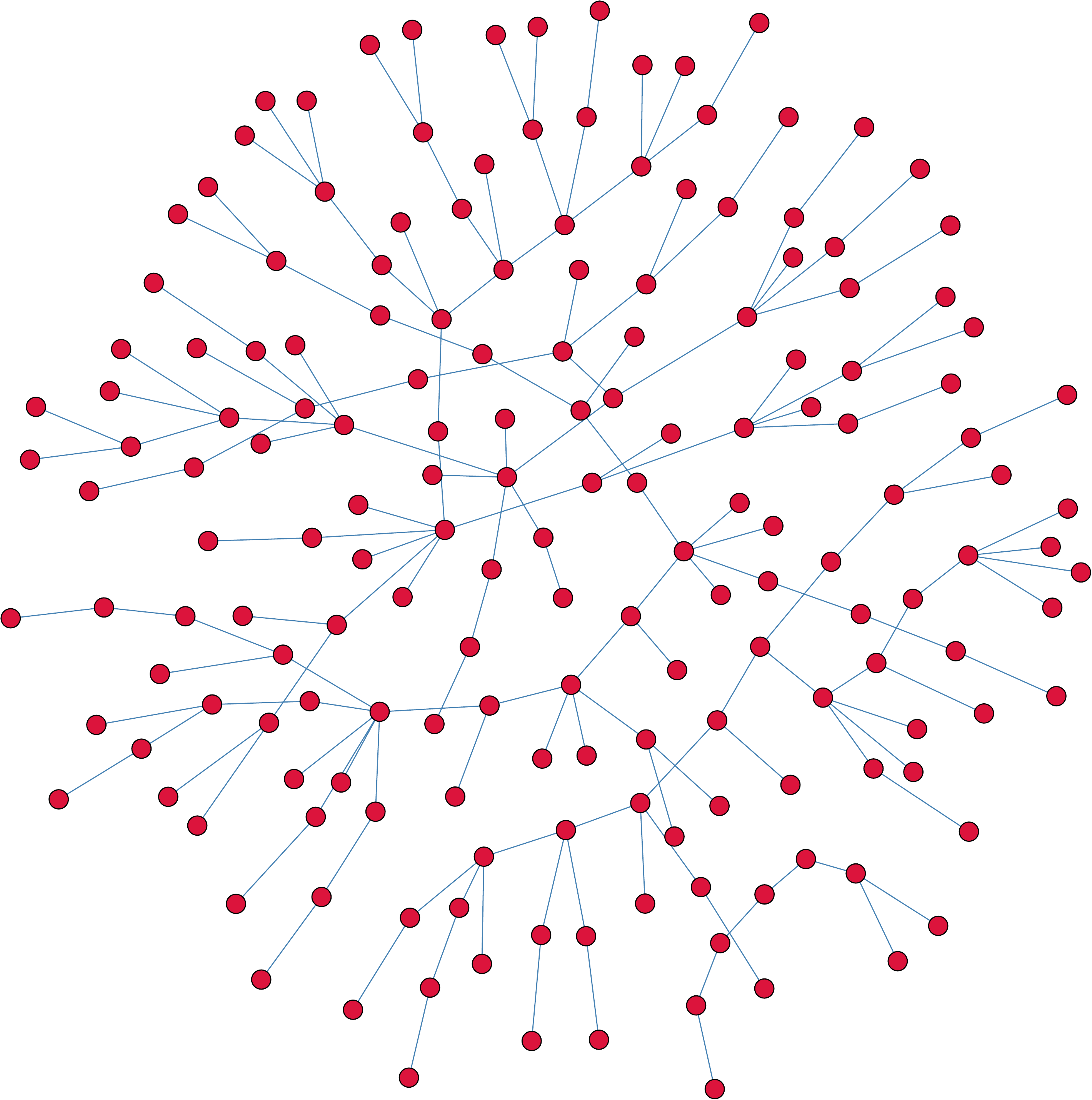}
\end{center}
\caption{\small\sf A five component forest with 174 vertices with a gravitational force.  
The figure on the left has gravity scaled up, while the figure on the right does not. }
\label{fig-forest5-50-gravD}
\end{figure}

In each case, the use of social gravity and scaling implies
that the resulting drawing tends to have vertices 
placed uniformly in a disk with few if any crossings. 
In addition, the algorithm tends to place the 
larger trees in a forest near the center of the disk.
(See Figures~\ref{fig-forest5-20-degGrav}
and~\ref{fig-forest5-50-gravD}.)

\subsection{The Impact of Scaling}
The impact of the scaling of social-gravity
forces on vertices over the course of the algorithm 
becomes important for the ultimate placement of 
vertices in a quality drawing.  For instance, examine the five component forest in the far left of Figure~\ref{fig-forest5-20-degGrav}.  
These trees have been drawn using a traditional force-directed algorithm. 
Components have each achieved a low energy state, and then begun to spread apart via repulsive forces.  
The same forest is then redrawn with gravity using two approaches.  
In the drawing in the center of Figure~\ref{fig-forest5-20-degGrav}, 
gravity is applied in increasing amounts, beginning with no gravity, and increasing according to a step function.  
In the drawing on the right, gravity is applied at a uniformly high level throughout the drawing.  
In drawings with low number of vertices, the quality of the drawing is robust to the method of gravity application.  
If we examine the larger forest in Figure~\ref{fig-forest5-50-gravD}, 
for example, with the left drawing having scaled gravity and the right drawing
having uniformly high gravity, we see that scaling gravity dramatically reduces
crossings in the final output.  
Our findings indicate that the benefits of using social gravity and scaling
are achieved as long as the algorithm begins with weak or no gravity, and increases, even with few steps, to the maximum gravitational force.

\subsection{The Impact of Using Different Forms of Centrality}
While the various forms of centrality represent related ideas, the variation in definition means that gravitational forces based on these distinct forms will produce differing results. This can be illustrated with high degree vertices that occur near edge of the graph in a classic force-embedding.  These vertices will have high degree centrality, but because of their exterior structural location, they will often have low closeness and betweenness centrality.  In degree centrality gravitational embedding, these vertices will be attracted to the center with more force, while embeddings with other centralities will leave them on the exterior of the graph.

In Figure~\ref{fig-t70} we see a comparison of the effects of gravity defined using different centralities on a tree of 70 vertices.  The top row is the tree embedded with the force-directed method with no gravitational force, and colored from left to right with degree, closeness, and betweenness centrality.  Red coloring represents vertices of high centrality, blue vertices have low centrality, and vertices in between are colored with a spectrum from red to blue according to their relative centrality in the graph.  The lower row consists of the results of the gravitational force-directed algorithm where the mass relates the centrality of the vertices, once again, from left to right with degree, closeness, and betweenness centrality.  

\section{Social Networks}\label{sec:soc-net}

\begin{figure}[hbt!]
\begin{center}
\includegraphics[width=.3\textwidth]{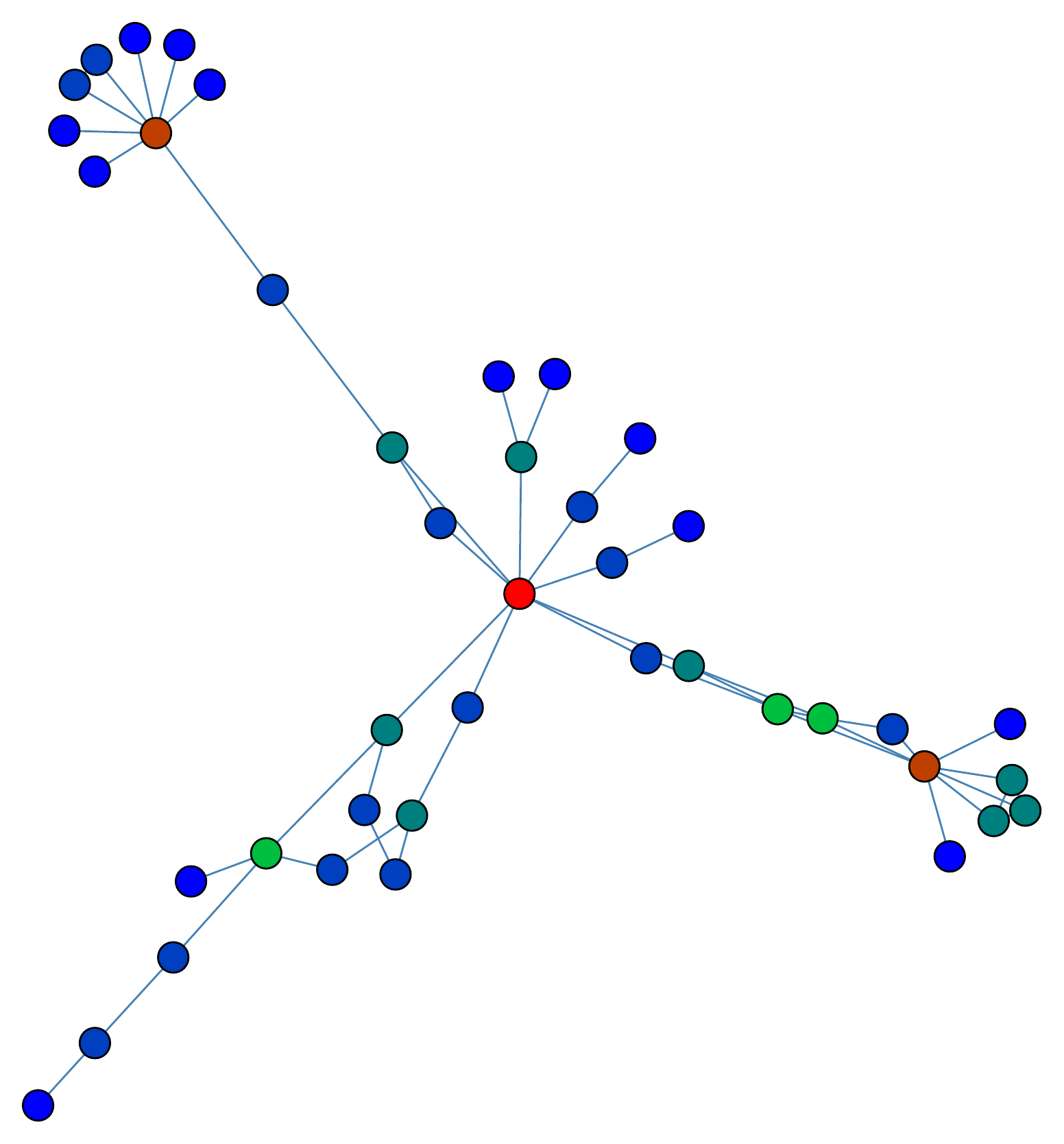}
\quad
\includegraphics[width=.3\textwidth]{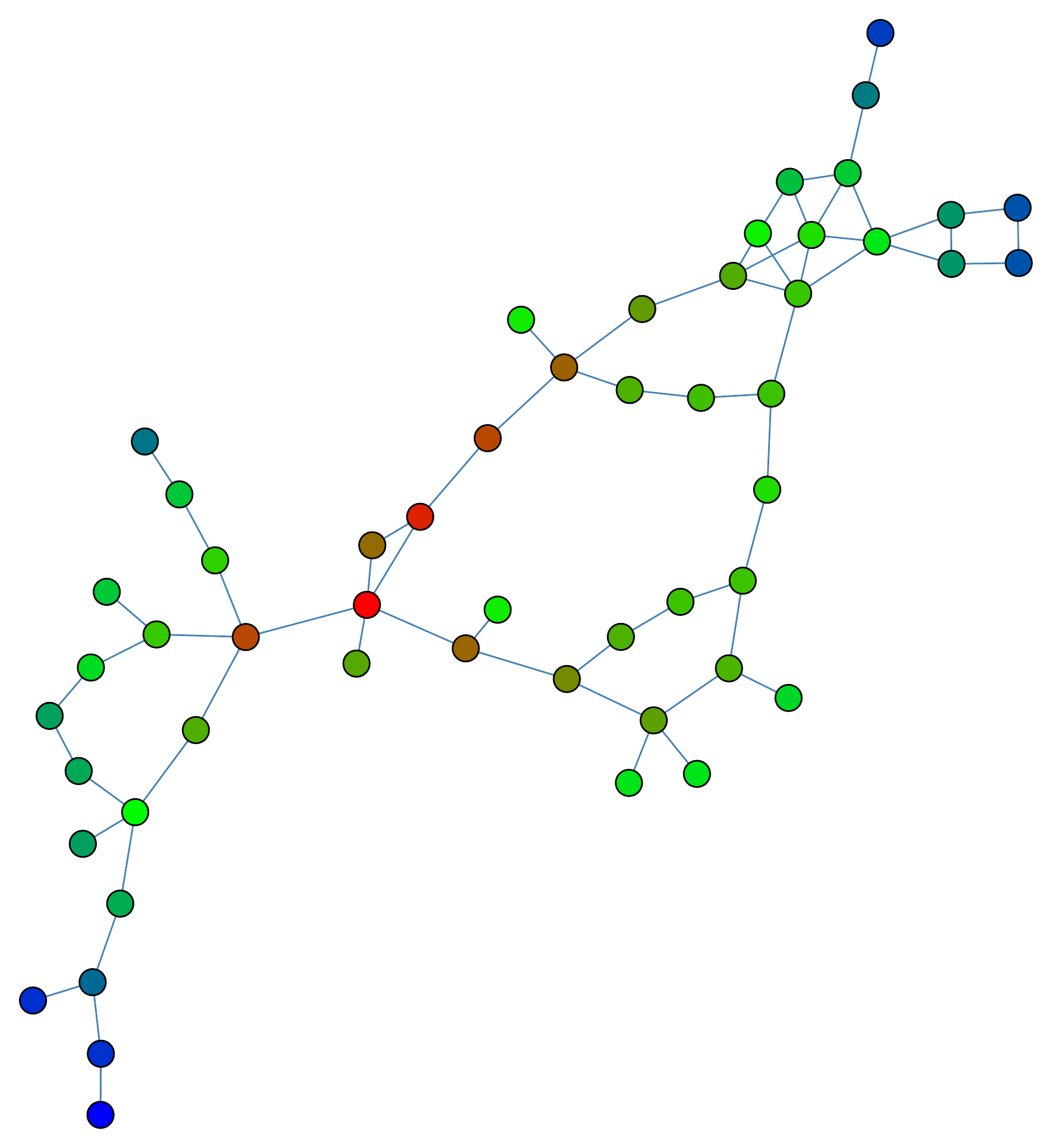}
\quad
\includegraphics[width=.3\textwidth]{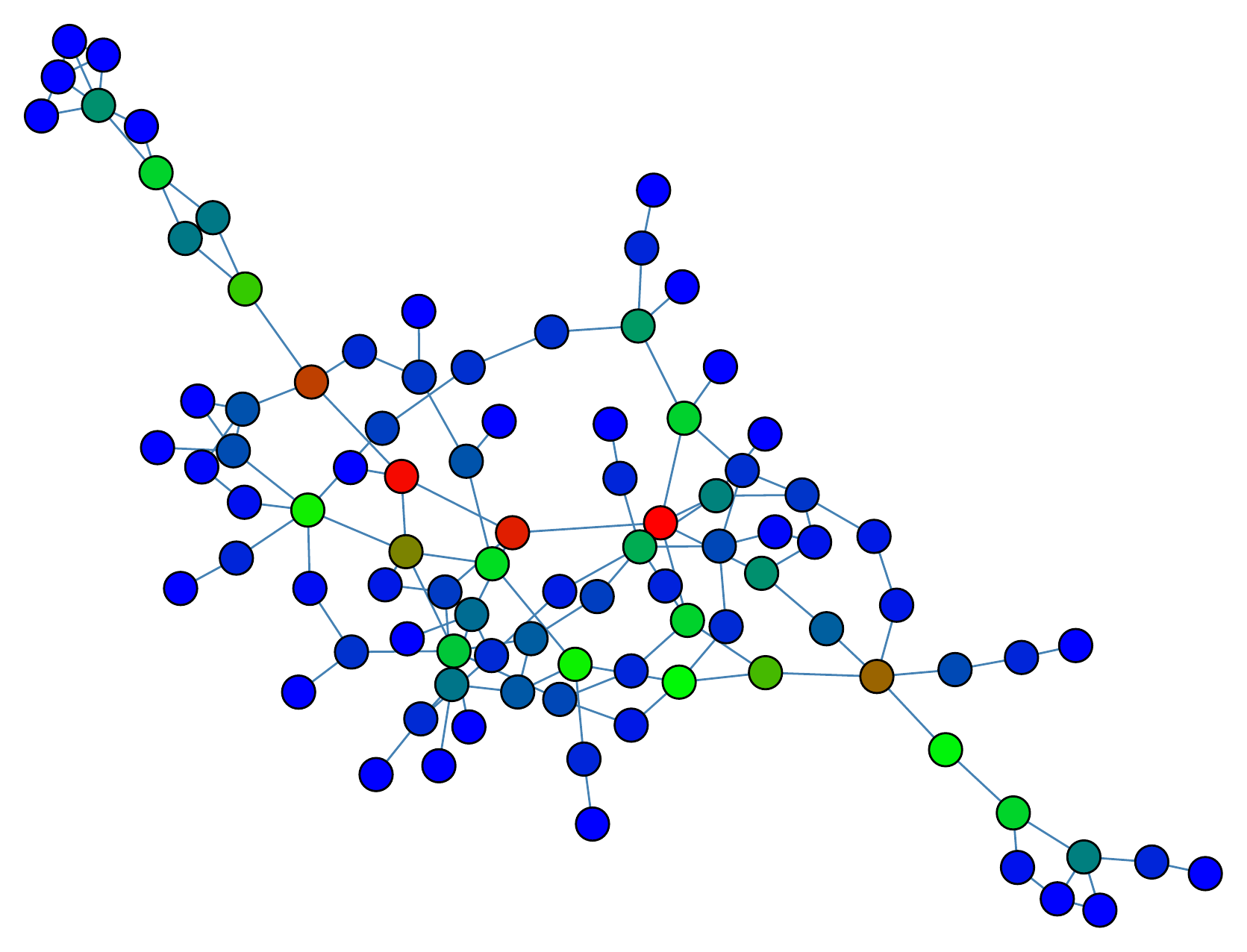}

\vspace{2em}

\includegraphics[width=.3\textwidth]{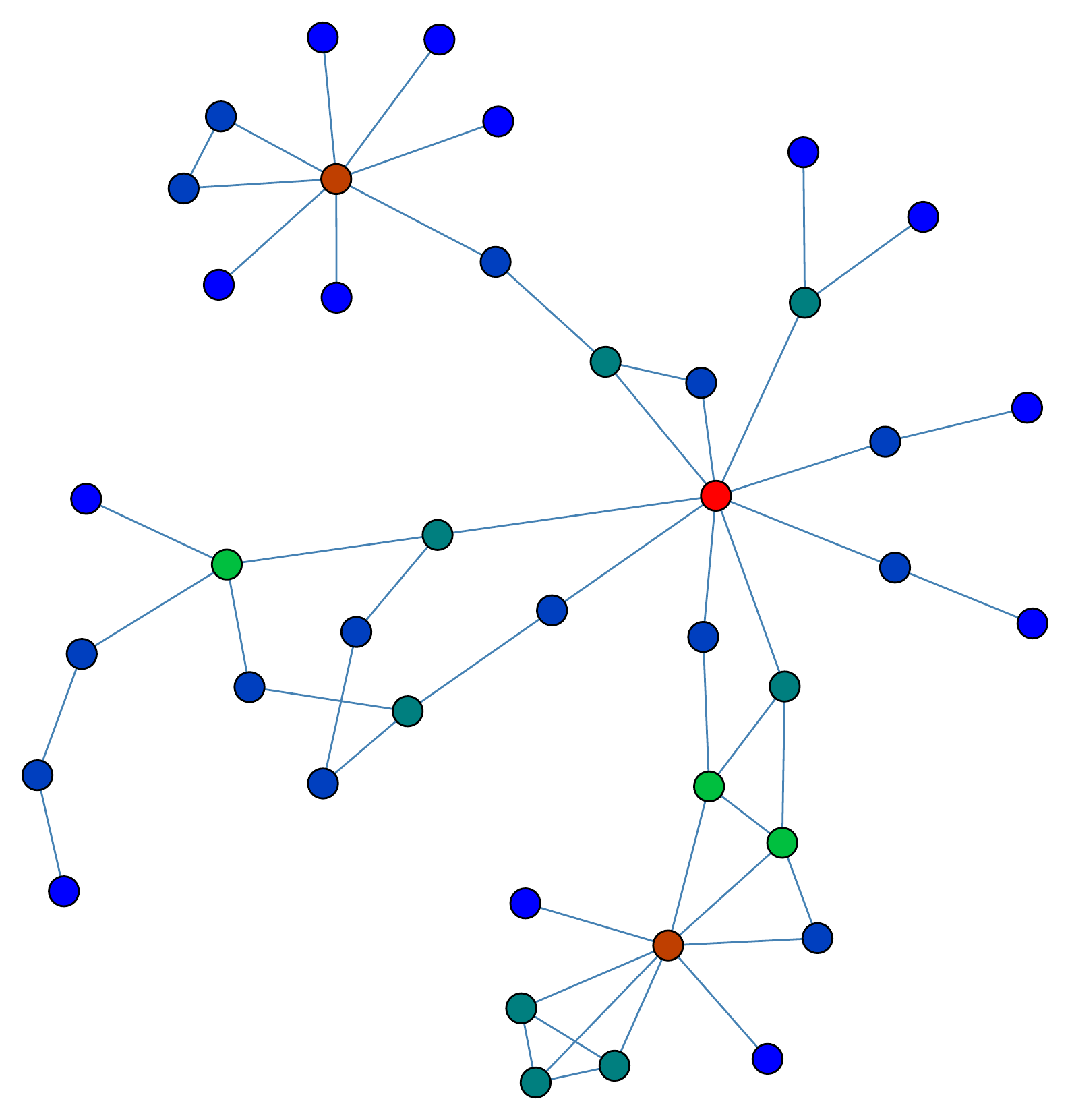}
\quad
\includegraphics[width=.3\textwidth]{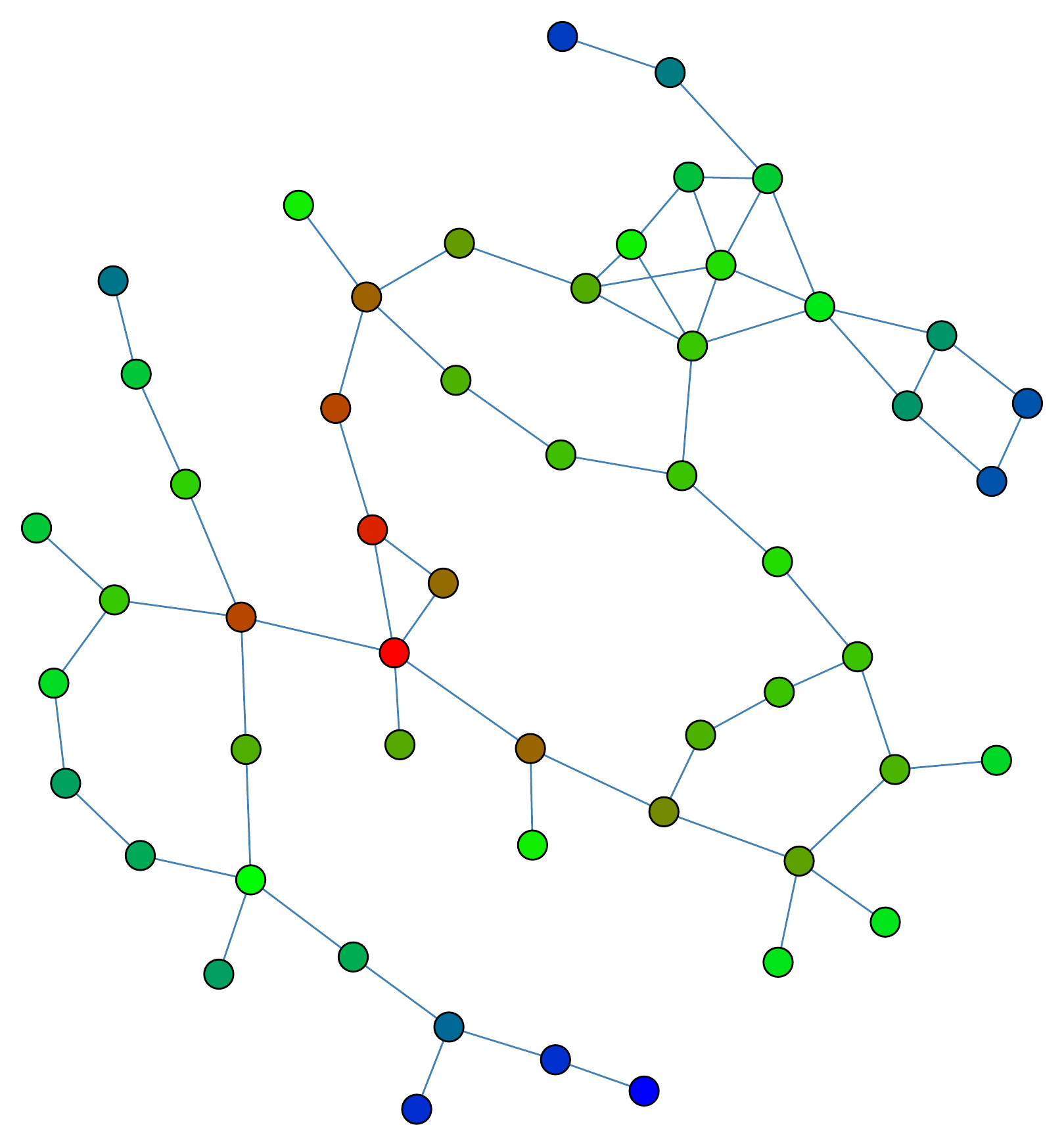}
\quad
\includegraphics[width=.3\textwidth]{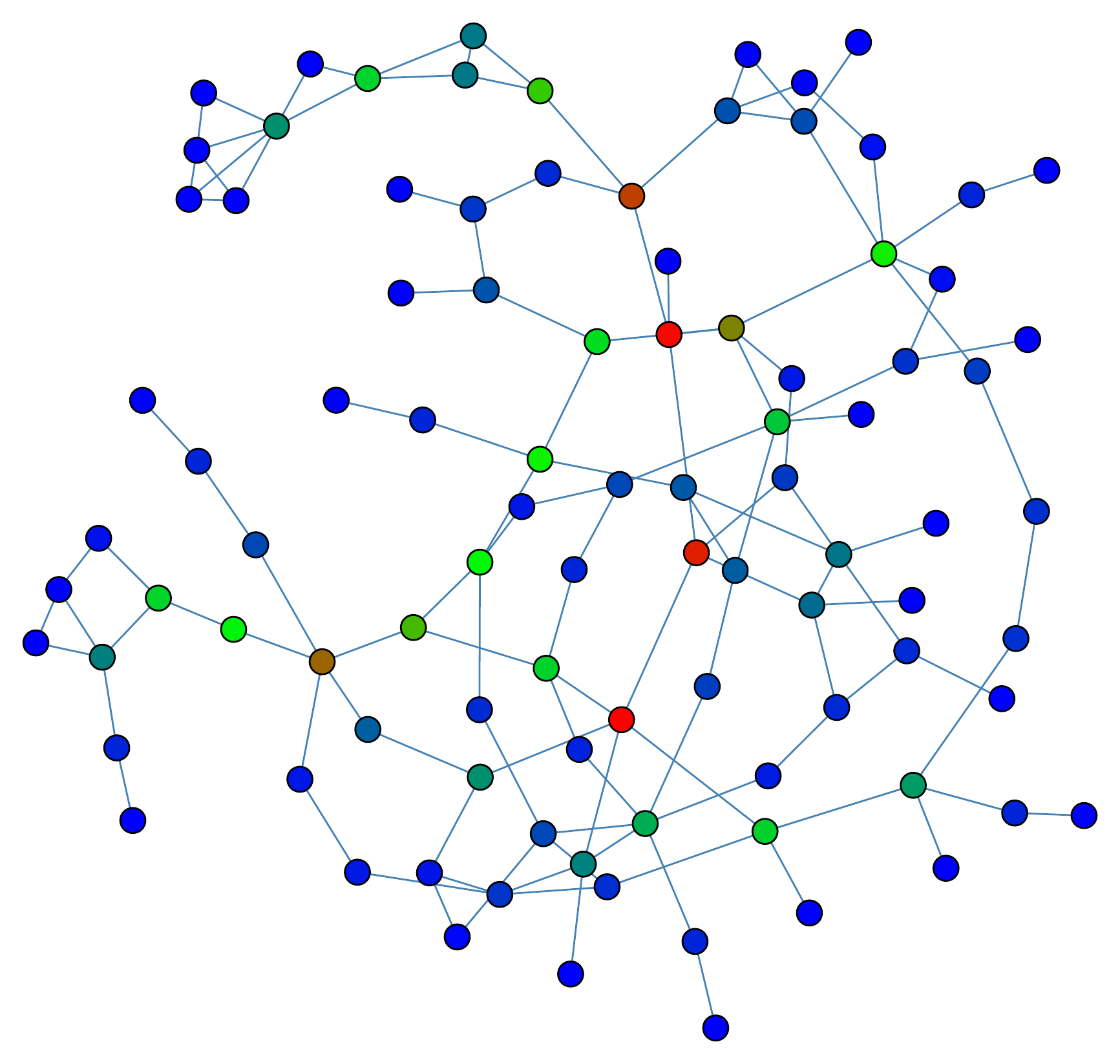}
\end{center}
\caption{\small\sf Comparison of graphs embedded with the classic force-directed algorithm(top row) and the gravitational force-directed algorithm(bottom row).   The columns from left to right are colored and utilize the gravitational force on degree, closeness, and betweenness centrality respectively. }
\label{fig-compGraphs}
\end{figure}

As our algorithm was motivated by the visualization of social networks, we chose example graphs in this application area as test inputs for it.
Figure~\ref{fig-compGraphs} shows several such graphs and their corresponding
gravitational drawings.
As the gravitational forces are exerted on vertices with high centrality, 
these are pulled towards the center of the drawing.  
This makes vertices that are more central to the social structure of 
the network visually localized.  
Prominent social forces are thus more easily identified, 
increasing discovery and communication of these social actors.

The overall effects of our algorithm on these graph drawings are similar to its effects on trees.
The drawings produced by our algorithm tend to be more space efficient, as
observed above.
In the classic force directed model, long components that are distant from the most central vertices of the graph are repulsed by many other vertices, with no compensatory attractive forces, and tend to be stretched out from the center, causing the drawing to have a large bounding box.
Since the gravitational force of our algorithm attracts these groupings of 
vertices, they are brought closer to the core of the graph, 
and are able to repulse each other, increasing visibility.  
This can been seen especially in the degree and betweenness 
centrality embeddings in Figure~\ref{fig-compGraphs}.

A less obvious trait of the gravitational force is that our drawings 
tend to exhibit is superior angular resolution than their 
classic force-directed counterparts.  
Vertices are distributed more uniformly as gravitational forces increase, 
and so adjacent vertices tend away from co-linearity.  
This phenomenon is clearly visible in the upper-left corner of the 
degree centrality drawings of Figure~\ref{fig-compGraphs}.  

As with all force-directed drawing algorithms, 
finding the appropriate balance of the competing forces is important.  
Our implementation, while certainly using a gravitational force that 
was strong enough to effect the final outcome of the algorithm, 
did not make it overpowering in comparison with the classic forces.  
This was done to preserve all the benefits exhibited by the classical approach, 
while trying to enhance very successful drawings in tangible ways.  
Obviously, if some characteristic is especially desired, 
like the central movement of vertices with high centrality, 
then one can modify the related forces accordingly.
We demonstrate our algorithm on several additional examples of social networks in an
appendix. 

\subsection{Lombardi-style Graph Drawing with Social Gravity}
\begin{figure}[hbt!]
\begin{center}
\includegraphics[width=.3\textwidth]{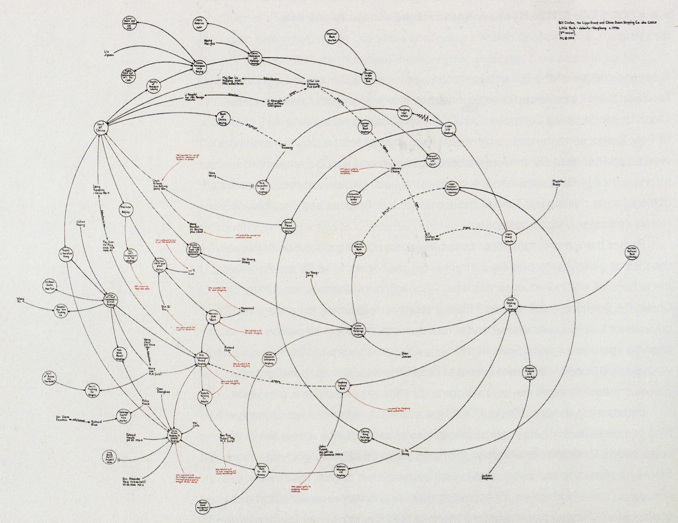}
\hspace{1em}
\includegraphics[width=.3\textwidth]{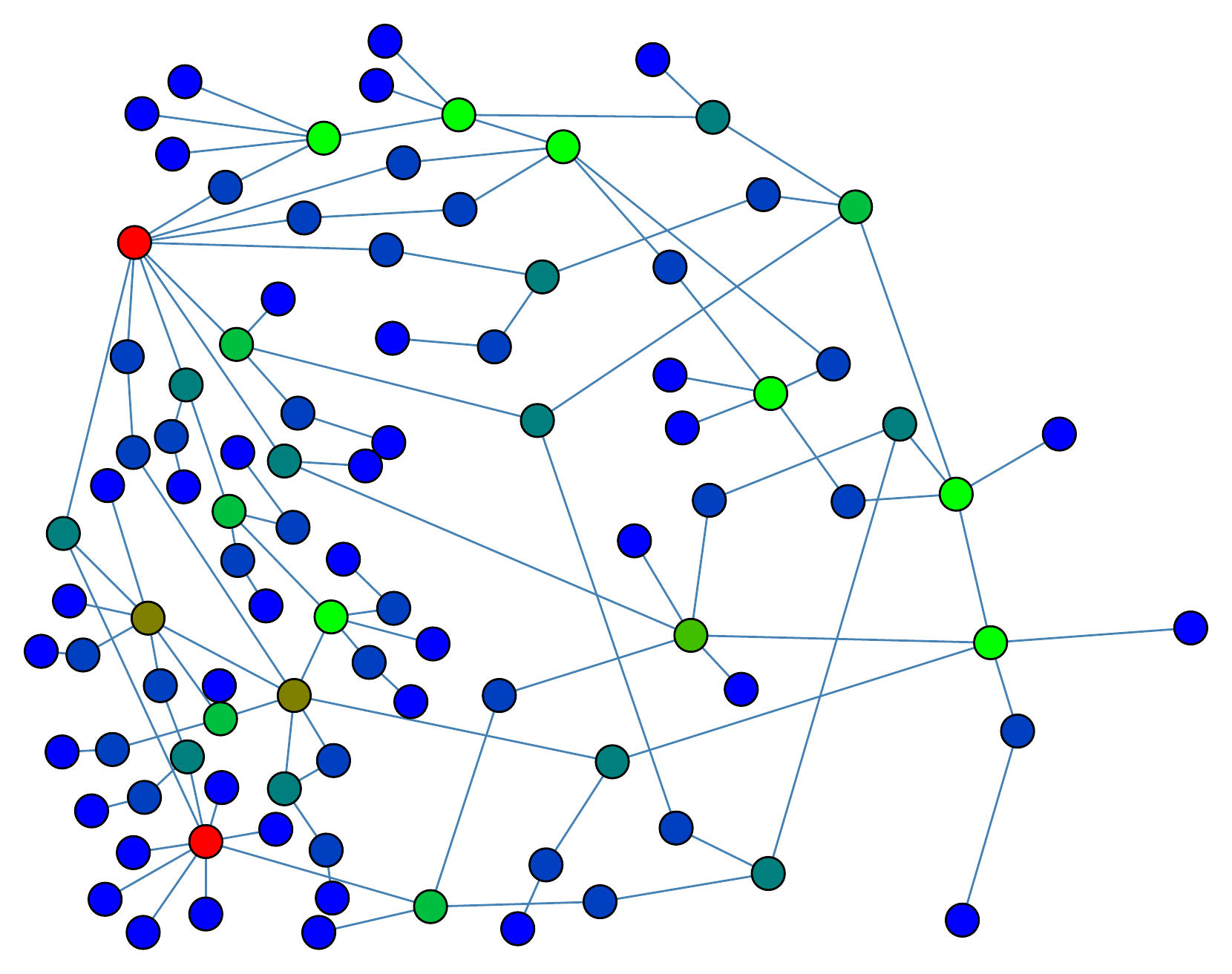}
\hspace{1em}
\includegraphics[width=.3\textwidth]{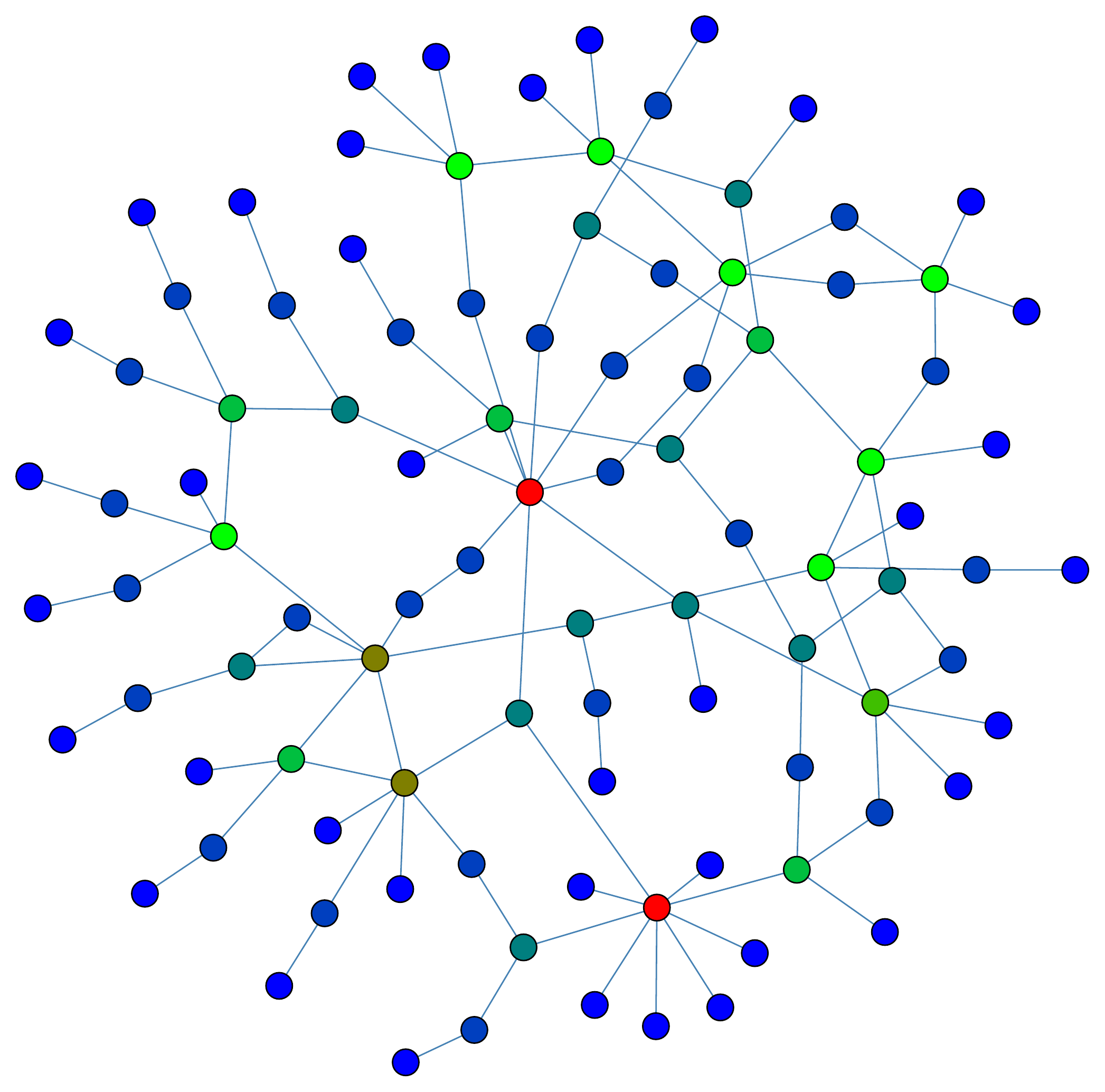}

\end{center}
\caption{\small\sf Lombardi's ``Bill Clinton, the Lippo Group, and China Ocean Shipping" network with straight-line approximation and degree centrality gravitational embedding. }
\label{fig-lombardi-lippo}
\end{figure}

\begin{figure}[hbt!]
\begin{center}
\includegraphics[width=.33\textwidth]{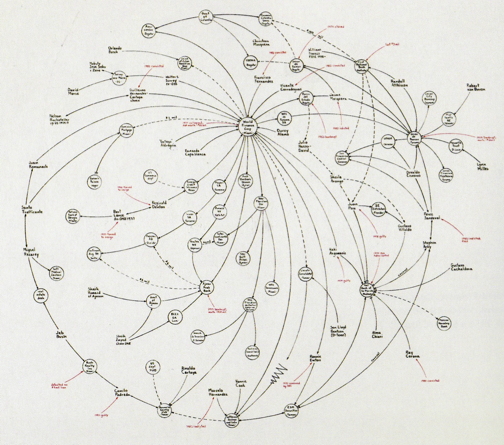}
\includegraphics[width=.3\textwidth]{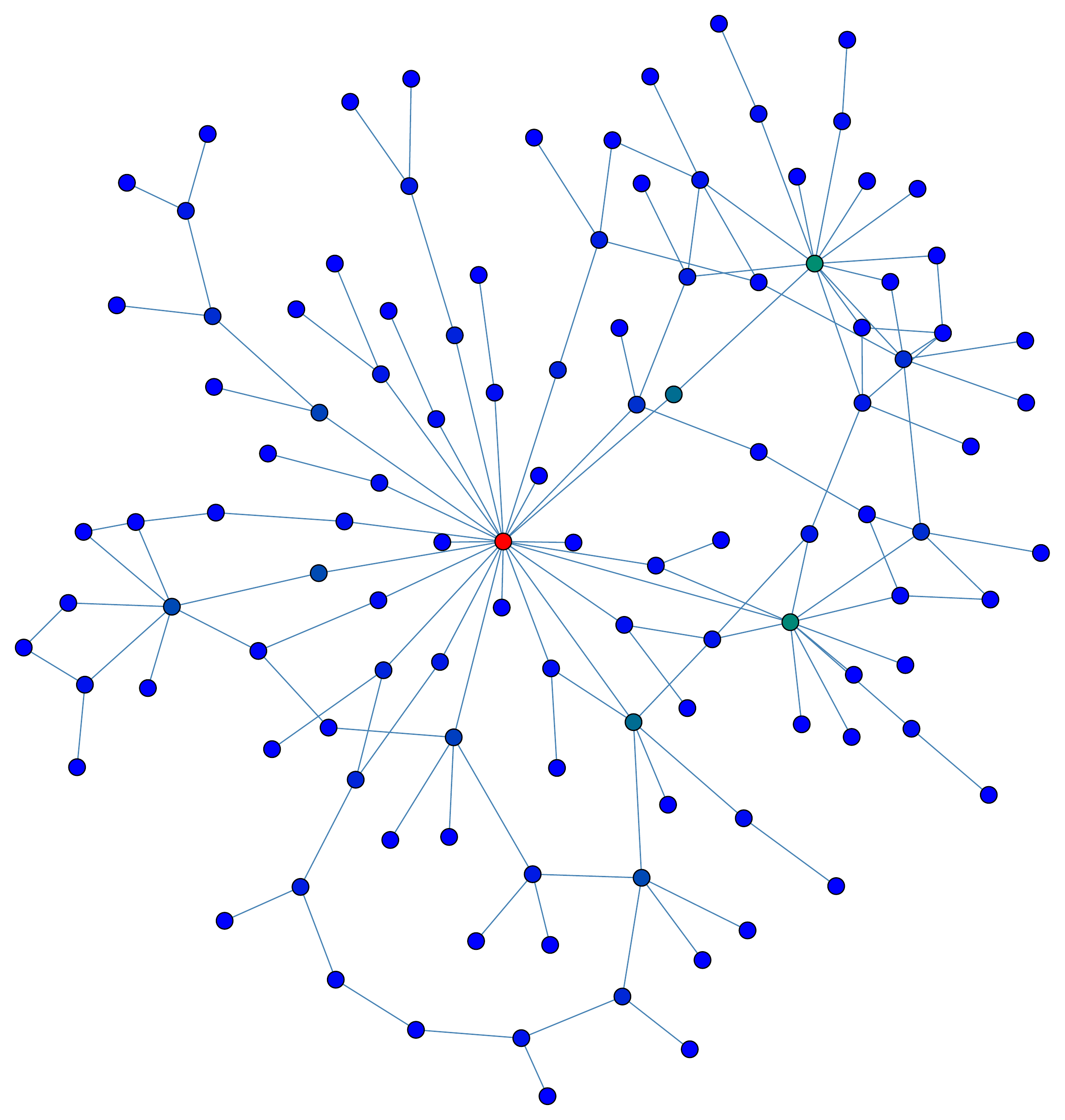}
\hspace{1em}
\includegraphics[width=.3\textwidth]{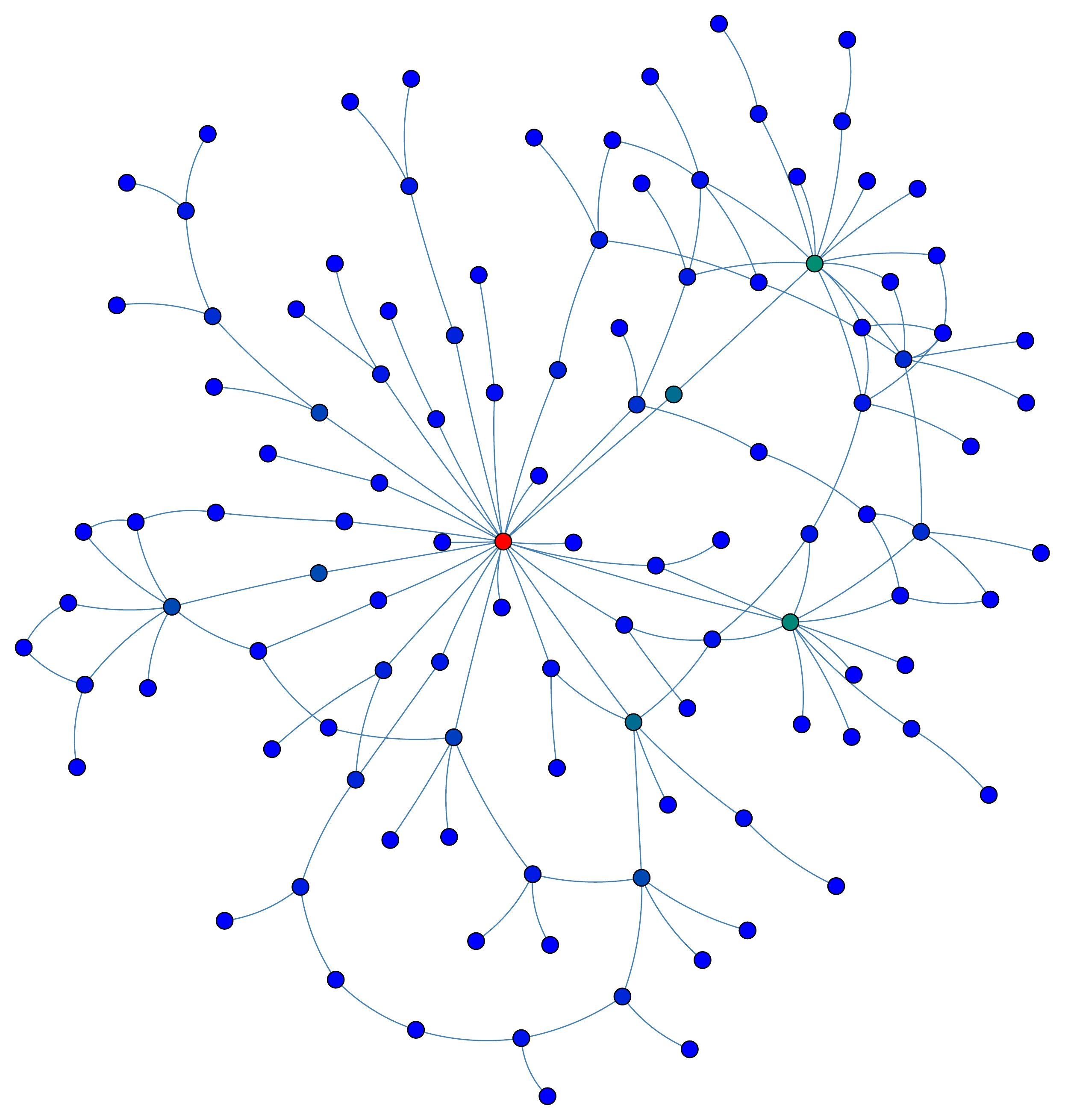}

\end{center}
\caption{\small\sf Lombardi's ``World Finance Corporation, Miami, Florida.''
From left to right: Lombardi's drawing, the betweenness centrality
gravitational drawing, and the central drawing with the addition of
force-directed Lombardi-style edges. }
\label{fig-lombardi-miami}
\end{figure}

We also can apply our approach to Lombardi's social networks
and Lombardi-style graph drawing.
Specifically,
in Figures~\ref{fig-lombardi-lippo}~and~\ref{fig-lombardi-miami} we have drawn two of the conspiracy networks documented by Mark Lombardi.  
In these graphs vertices represent people or business entities and edges represent business dealings. 
We illustrate the drawings of these networks with our gravitational force-directed algorithm.  
In Figure ~\ref{fig-lombardi-miami} we show our method in conjunction with the 
force-directed approach to Lombardi-style 
drawing~\cite{ccgkt-fdlgd-12}, which 
uses circular-arc edges.  

In the left panel of Figure~\ref{fig-lombardi-lippo} 
we see Lombardi's ``Bill Clinton, the Lippo Group, and China Ocean Shipping'' 
graph.  
This graph highlights the possible associations between illegal arms 
dealing by Chinese nationalists in Los Angeles and 
supposed White House campaign-finance corruption~\cite{lh-mlgn-03}.  
The middle panel is a straight line approximation of Lombardi's drawing 
colored with degree centrality.  
Note that many of the high degree vertices are condensed in the lower left section of the drawing.  
It is also clear that the nodes are packed much more densely in this region.  
If we examine the right panel, we see that the high degree vertices, 
plausibly the actors more central to the presented power relationship, 
are now more centralized.  
As we would expect, vertices also have a much more uniform spacial distribution.
Note the increase in angular resolution of the central high centrality vertex, and the vertex in the upper left of the central drawing. 

The left of Figure~\ref{fig-lombardi-miami} contains another of 
Lombardi's drawings,``World Finance Corporation, Miami, Florida".  
This network represents the plausible role played by WFC 
in trafficking Colombian drugs and money laundering of the profits.  
The central drawing is the network embedded with betweenness centrality gravitational forces.  
As is desirable, the vertex of highest centrality has been centered in the output.
The right drawing is the result of the application of the dummy-vertex approach to force-directed Lombardi-style drawing to the result of the our gravitational algorithm.  While not as consistent as Lombardi's hand drawing, this automated approach does succeed in replicating some of the long arcs through multiple nodes found in the original.  The creation of circular arc edges is done with classical forces, and thus augmenting the implementation of the gravitational algorithm to create Lombardi-style drawings is very simple.

\section{Conclusions and Directions for Future Work}

The addition of social gravity 
to force-directed drawings produces several positive results.
Vertices with high centrality become centrally located, vertices are distributed more uniformly in the drawing, and drawn graphs exhibit  superior angular resolution.  
Moreover, scaling gravity has proven to be important to ensuring 
these positive traits.  
In particular,
the benefits of social gravity and scaling lead to drawings of trees and 
forests that are aesthetically pleasing, compact, tend to be crossing free, and have uniform edge length.  
Social networks also exhibit these positive traits, although in some cases, 
this produces crossings that would have been otherwise avoided.  

There are several directions for future work.
Study of the ideal balance between gravitational forces and classical forces is left to future work.  
We have also shown that the force-directed approach to Lombardi-style drawing can be used in conjunction with our algorithm.  
We recommend future study of how the Lombardi forces should interact with the differentiated final edge length created by the output of the 
gravitational algorithm. 
Finally, there is also room for improvement in the runtime of our algorithm, which is quadratic as implemented, using methods from $n$-body simulation~\cite{Hu-TMJ-05}.

\subsection*{Acknowledgements}
The work on this paper was supported in part by the U.S. Office of Naval
Research, under MURI grant N00014-08-1-1015, and the U.S. National
Science Foundation, under grants 0830403, 0953071, and 1011840.

{\raggedright
\bibliographystyle{abuser}
\bibliography{atomic-lombardi,lombardi,stephen}

\begin{thebibliography}{10}

\bibitem{BorMehBra-09}
S.~P. Borgatti, A.~Mehra, D.~J. Brass, and G.~Labianca.
\newblock {Network Analysis in the Social Sciences}.
\newblock {\em Science} 323(5916):892{--}895, 2009,
  \href{http://dx.doi.org/10.1126/science.1165821}%
{doi:10.1126/science.1165821}.

\bibitem{BraHimRoh-GD-96}
F.~Brandenburg, M.~Himsolt, and C.~Rohrer.
\newblock {An experimental comparison of force-directed and randomized graph
  drawing algorithms}.
\newblock {\em 3rd Symp. on Graph Drawing}, pp.~76{--}87. Springer, LNCS 1027,
  1996, \href{http://dx.doi.org/10.1007/BFb0021792}%
{doi:10.1007/BFb0021792}.

\bibitem{Bra-JMS-01}
U.~Brandes.
\newblock {A faster algorithm for betweenness centrality}.
\newblock {\em The Journal of Mathematical Sociology} 25(2):163{--}177, 2001,
  \href{http://dx.doi.org/10.1080/0022250X.2001.9990249}%
{doi:10.1080/0022250X.2001.9990249}.

\bibitem{Bra-DG-05}
U.~Brandes.
\newblock {Drawing on Physical Analogies}.
\newblock {\em Drawing Graphs}, pp.~71{--}86. Springer, LNCS 2025, 2001.

\bibitem{bfw-sn-13}
U.~Brandes, L.~C. Freeman, and D.~Wagner.
\newblock Social networks.
\newblock {\em {Handbook of Graph Drawing and Visualization}}, pp.~26-1--26-32.
  Chapman {\&} Hall/CRC, 2013.

\bibitem{BraKenRaa-M-06}
U.~Brandes, P.~Kenis, and J.~Raab.
\newblock {Explanation Through Network Visualization}.
\newblock {\em Methodology: European Journal of Research Methods for the
  Behavioral and Social Sciences} 2(1):16{--}23, 2006,
  \href{http://dx.doi.org/10.1027/1614-2241.2.1.16}%
{doi:10.1027/1614-2241.2.1.16}.

\bibitem{BraKenRaa-JTP-99}
U.~Brandes, P.~Kenis, J.~Raab, V.~Schneider, and D.~Wagner.
\newblock {Explorations into the Visualization of Policy Networks}.
\newblock {\em Journal of Theoretical Politics} 11(1):75{--}106, 1999,
  \href{http://dx.doi.org/10.1177/0951692899011001004}%
{doi:10.1177/0951692899011001004}.

\bibitem{BraPic-JGAA-11}
U.~Brandes and C.~Pich.
\newblock {More Flexible Radial Layout}.
\newblock {\em Journal of Graph Algorithms and Applications} 15(1):157{--}173,
  2011, \url{http://jgaa.info/accepted/2011/BrandesPich2011.15.1.pdf}.

\bibitem{Fre-MMSNA-05}
P.~J. Carrington, J.~Scott, and S.~Wasserman.
\newblock {\em {Models and Methods in Social Network Analysis}}.
\newblock Cambridge University Press, 2005, p.~248{\^O}{\c{C}}{\^o}269.
\newblock Graphic techniques for exploring social network data. by Linton C.
  Freeman.

\bibitem{ccgkt-fdlgd-12}
R.~Chernobelskiy, K.~Cunningham, M.~T. Goodrich, S.~Kobourov, and L.~Trott.
\newblock Force-directed {Lombardi}-style graph drawing.
\newblock {\em Graph Drawing}, pp.~320--331. Springer, LNCS 7034, 2012,
  \url{http://dx.doi.org/10.1007/978-3-642-25878-7_31}.

\bibitem{ccm-vrsn-12}
C.~Correa, T.~Crnovrsanin, and K.-L. Ma.
\newblock Visual reasoning about social networks using centrality sensitivity.
\newblock {\em IEEE Trans. on Visualization and Computer Graphics}
  18(1):106--120, jan. 2012, \href{http://dx.doi.org/10.1109/TVCG.2010.260}%
{doi:10.1109/TVCG.2010.260}.

\bibitem{DeSinWon-STI-04}
P.~De, A.~E. Singh, T.~Wong, W.~Yacoub, and A.~M. Jolly.
\newblock {Sexual network analysis of a gonorrhoea outbreak}.
\newblock {\em Sexually transmitted infections} 80(4):280{--}5, August 2004,
  \href{http://dx.doi.org/10.1136/sti.2003.007187}%
{doi:10.1136/sti.2003.007187}.

\bibitem{BatEadTam-98}
G.~Di~Battista, P.~Eades, R.~Tamassia, and I.~G. Tollis.
\newblock {\em {Graph Drawing: Algorithms for the Visualization of Graphs}}.
\newblock Prentice Hall PTR, Upper Saddle River, NJ, USA, 1998.

\bibitem{degkl-ppld-12}
C.~Duncan, D.~Eppstein, M.~T. Goodrich, S.~Kobourov, and M.~L{\"o}ffler.
\newblock Planar and poly-arc {Lombardi} drawings.
\newblock {\em Graph Drawing}, pp.~308--319. Springer, LNCS 7034, 2012,
  \url{http://dx.doi.org/10.1007/978-3-642-25878-7_30}.

\bibitem{degkn-dtwpa-10}
C.~A. Duncan, D.~Eppstein, M.~T. Goodrich, S.~G. Kobourov, and
  M.~N{\"o}llenburg.
\newblock {Drawing Trees with Perfect Angular Resolution and Polynomial Area}.
\newblock {\em 18th Symp. on Graph Drawing}, pp.~183{--}194, 2010.

\bibitem{degkn-ldg-10}
C.~A. Duncan, D.~Eppstein, M.~T. Goodrich, S.~G. Kobourov, and
  M.~N{\"o}llenburg.
\newblock {Lombardi} drawings of graphs.
\newblock {\em 18th Symp. on Graph Drawing}, pp.~195{--}207, 2010.

\bibitem{de-vfmfg-02}
T.~Dwyer and P.~Eades.
\newblock Visualising a fund manager flow graph with columns and worms.
\newblock {\em 6th Int. Conf. on Information Visualisation}, pp.~147--152,
  2002, \href{http://dx.doi.org/10.1109/IV.2002.1028770}%
{doi:10.1109/IV.2002.1028770}.

\bibitem{Fre-SN-78}
L.~C. Freeman.
\newblock {Centrality in social networks conceptual clarification}.
\newblock {\em Social Networks} 1(3):215{--}239, 1979,
  \href{http://dx.doi.org/10.1016/0378-8733(78)90021-7}%
{doi:10.1016/0378-8733(78)90021-7}.

\bibitem{Fre-JSS-00}
L.~C. Freeman.
\newblock {Visualizing social networks}.
\newblock {\em Journal of Social Structure} 1(1), 2000.

\bibitem{FriLudMeh-GD-94}
A.~Frick, A.~Ludwig, and H.~Mehldau.
\newblock {A Fast Adaptive Layout Algorithm for Undirected Graphs}.
\newblock {\em Graph Drawing}, pp.~388{--}403. Springer Berlin / Heidelberg,
  Lecture Notes in Computer Science 1984, 1994.

\bibitem{FruRei-SPE-91}
T.~M.~J. Fruchterman and E.~M. Reingold.
\newblock {Graph drawing by force-directed placement}.
\newblock {\em Software: Practice and Experience} 21(11):1129{--}1164, 1991,
  \href{http://dx.doi.org/10.1002/spe.4380211102}%
{doi:10.1002/spe.4380211102}.

\bibitem{Fur-10}
B.~Furht.
\newblock {\em {Handbook of Social Network Technologies and Applications}}.
\newblock Springer-Verlag, 2010.

\bibitem{GajGooKob-CG-04}
P.~Gajer, M.~T. Goodrich, and S.~G. Kobourov.
\newblock {A multi-dimensional approach to force-directed layouts of large
  graphs}.
\newblock {\em Comp. Geometry: Theory and Applications} 29(1):3{--}18, 2004.

\bibitem{GajKob-GD-01}
P.~Gajer and S.~Kobourov.
\newblock {GRIP: Graph Drawing with Intelligent Placement}.
\newblock {\em Graph Drawing}, pp.~104{--}109. Springer, LNCS 1984, 2001.

\bibitem{Hu-TMJ-05}
Y.~F. Hu.
\newblock {Efficient and high quality force-directed graph drawing}.
\newblock {\em The Mathematica Journal} 10(1):37{--}71, 2005.

\bibitem{Klo-SN-81}
A.~S. Klovdahl.
\newblock {A note on images of networks}.
\newblock {\em Social Networks} 3:197--{\^O}{\c{C}}{\^o}214, 1981.

\bibitem{lh-mlgn-03}
M.~Lombardi and R.~Hobbs.
\newblock {\em {Mark Lombardi: Global Networks}}.
\newblock Independent Curators, 2003.

\bibitem{NewNew-07}
H.~Newman and J.~O. Newman.
\newblock {\em {A Genealogical Chart of Greek Mythology}}.
\newblock The University of North Carolina Press, 2007.

\bibitem{PotMutRot-STI-02}
J.~J. Potterat, S.~Q. Muth, R.~B. Rothenberg, H.~Zimmerman-Rogers, D.~L. Green,
  J.~E. Taylor, M.~S. Bonney, and H.~A. White.
\newblock {Sexual network structure as an indicator of epidemic phase}.
\newblock {\em Sexually transmitted infections} 78 Suppl 1:i152--8, April 2002.

\bibitem{PotPhiPlu-STI-02}
J.~J. Potterat, L.~Phillips-Plummer, S.~Q. Muth, R.~B. Rothenberg, D.~E.
  Woodhouse, T.~S. Maldonado-Long, H.~P. Zimmerman, and J.~B. Muth.
\newblock {Risk network structure in the early epidemic phase of HIV
  transmission in Colorado Springs}.
\newblock {\em Sexually transmitted infections} 78 Suppl 1:i159--63, April
  2002.

\bibitem{Sander1999175}
G.~Sander.
\newblock Graph layout for applications in compiler construction.
\newblock {\em Theoretical Computer Science} 217(2):175--214, 1999,
  \href{http://dx.doi.org/10.1016/S0304-3975(98)00270-9}%
{doi:10.1016/S0304-3975(98)00270-9},
  \url{http://www.sciencedirect.com/science/article/pii/S0304397598002709}.

\bibitem{SteZel-SN-89}
K.~Stephenson and M.~Zelen.
\newblock {Rethinking centrality: Methods and examples}.
\newblock {\em Social Networks} 11(1):1{--}37, 1989,
  \href{http://dx.doi.org/10.1016/0378-8733(89)90016-6}%
{doi:10.1016/0378-8733(89)90016-6}.

\end{thebibliography}
}

\newpage
\appendix
\section{Additional Drawings of Social Networks}
\begin{figure}[hbt!]
\vspace*{-10pt}
\begin{center}
\includegraphics[width=.425\textwidth]{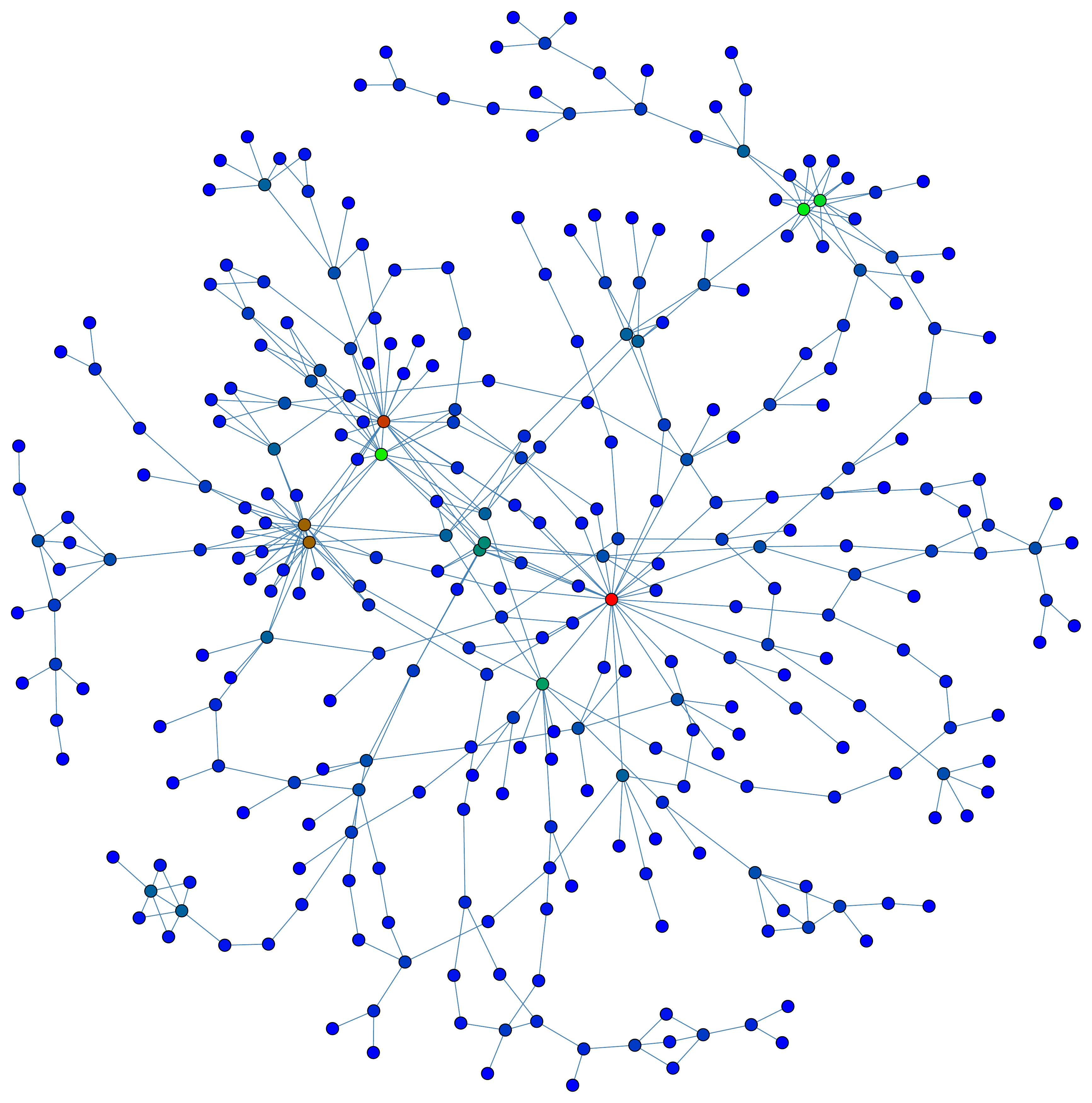}
\end{center}
\vspace*{-12pt}
\caption{\small\sf A genealogy graph of the Greek gods~\cite{NewNew-07}.}
\label{fig-greek}
\end{figure}

\begin{figure}[hbt!]
\begin{center}
\includegraphics[width=.425\textwidth]{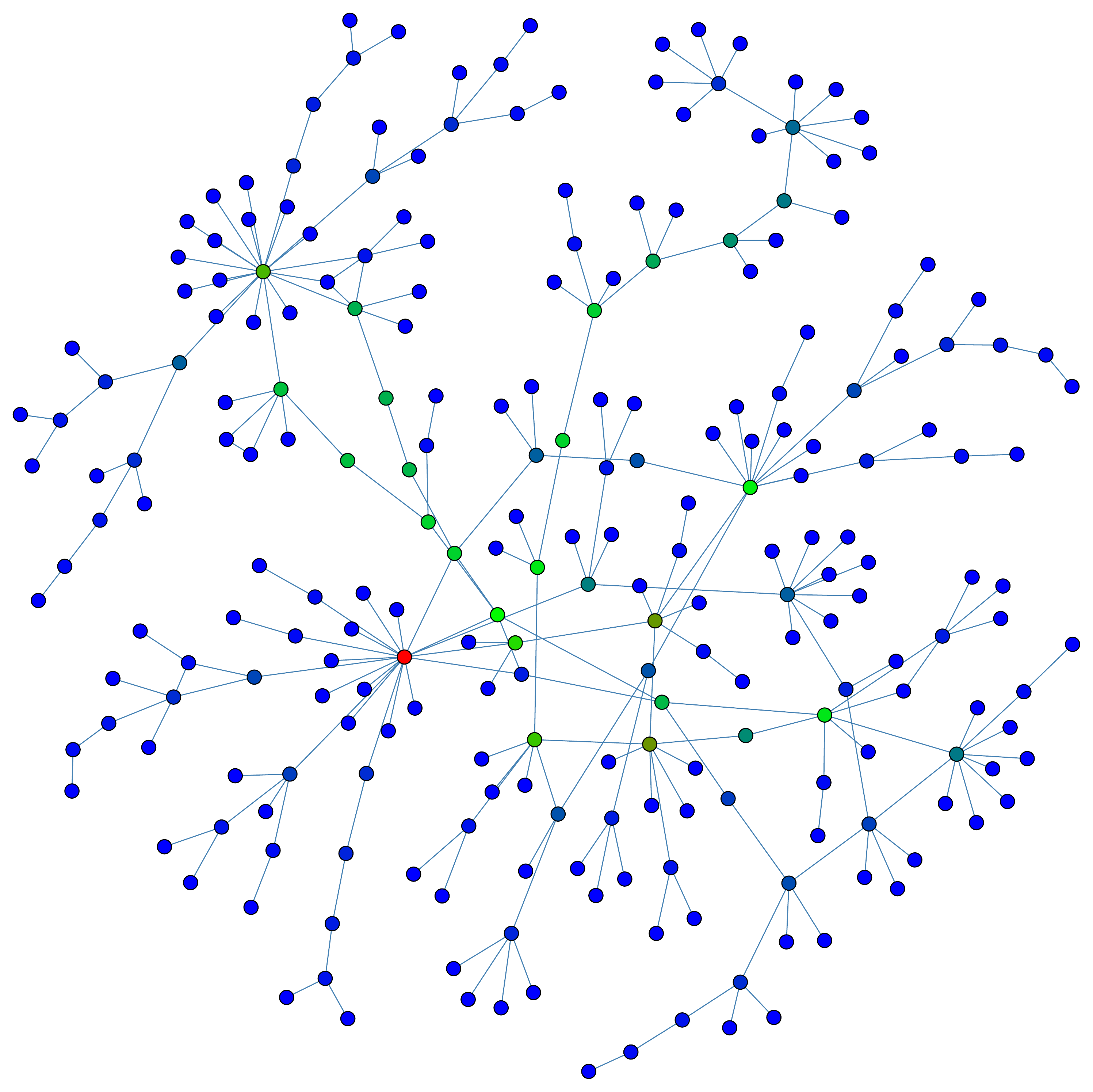}\hfill
\includegraphics[width=.425\textwidth]{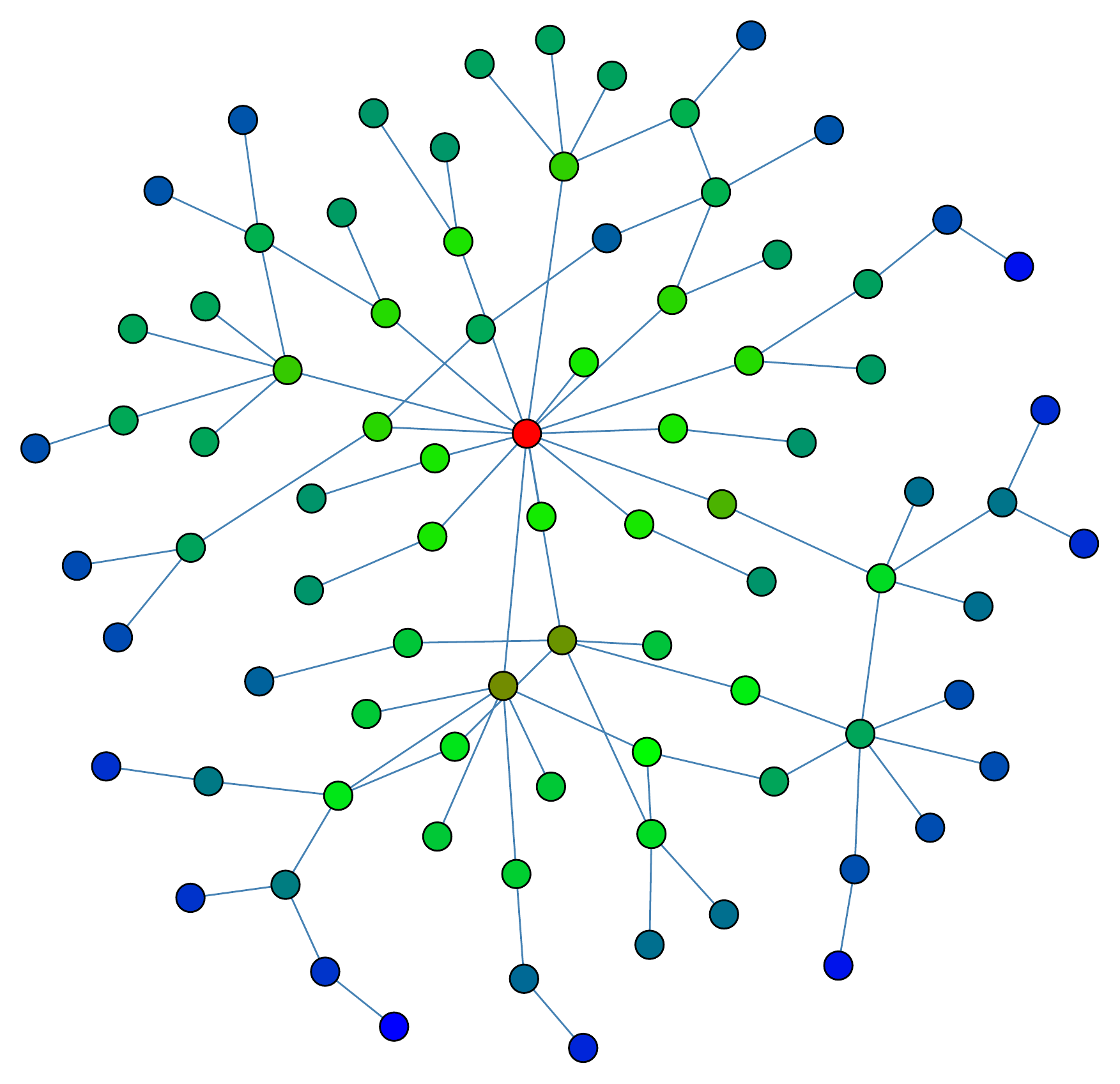}
\end{center}
\vspace*{-12pt}
\caption{\small\sf Sexual networks corresponding on the right to an HIV
outbreak~\cite{PotPhiPlu-STI-02}, and a gonorrhea outbreak on the
left~\cite{DeSinWon-STI-04, PotMutRot-STI-02}.}
\label{fig-hiv}
\label{fig-mst2}
\end{figure}

We demonstrate our algorithm on some additional types of social networks in this
appendix. 
In Figure~\ref{fig-greek} we draw a graph of genealogy of the Greek gods and related characters. 
The edges in this graph are drawn from parent to child. 
In an another type of social network,
we draw sexual networks collected by voluntary surveys during sexually 
transmitted infection outbreaks in Figure~\ref{fig-hiv}.  

\end{document}